\pdfoutput=1
\documentclass[journal]{IEEEtran}

\pdfpageattr{/Group << /S /Transparency /I true /CS /DeviceRGB >>}

\usepackage[utf8]{inputenc}
\usepackage[T1]{fontenc}
\usepackage{textcomp}
\usepackage{silence}
\usepackage{fixltx2e}
\usepackage{microtype}
\usepackage{calc}
\usepackage[normalem]{ulem}
\usepackage{balance}
\usepackage{lipsum}

\usepackage[range-phrase=--,per-mode=symbol-or-fraction,binary-units=true,range-units=single,list-units=single,detect-all]{siunitx}
\usepackage{silence}
\AtBeginDocument{
	\DeclareSIUnit\bit{bit}
	\DeclareSIUnit\byte{Byte}
	\DeclareSIUnit\decibeli{dBi}
	\DeclareSIUnit\decibelm{dBm}
	\DeclareSIUnit\mph{mph}
	\DeclareSIUnit\resourceblock{RB}
	\DeclareSIUnit\vehicle{veh}
	\DeclareSIUnit\watthour{Wh}
}

\usepackage{csquotes}
\usepackage[backend=biber,style=ieee,doi=false,isbn=false,mincitenames=1,maxcitenames=2]{biblatex}
\DeclareFieldFormat{sentencecase}{#1} % never apply sentence casing, even if bibtex field is unprotected
\DeclareFieldFormat{titlecase}{#1} % never apply title casing, even if bibtex field is unprotected
\usepackage{xpatch}
\xpatchbibmacro{textcite}{\addspace}{\addnbspace}{}{}
\xpatchbibmacro{Textcite}{\addspace}{\addnbspace}{}{}
\DefineBibliographyStrings{english}{
	andothers = et~al\adddot\addspace
}

\usepackage{amsmath}

\usepackage{amssymb}
\usepackage{amsfonts}
\usepackage{amsthm}

\usepackage{booktabs}
\usepackage{url}

\usepackage[inline]{enumitem}
\usepackage{subfigure}
\usepackage[pdftex]{graphicx}
\DeclareGraphicsExtensions{.pdf,.png,.jpg,.tikz}
\usepackage{tikz}
\usetikzlibrary{arrows}
\usetikzlibrary{calc}
\usetikzlibrary{chains}
\usetikzlibrary{scopes}

\usepackage[american]{babel}
\usepackage{hyphenat}

\usepackage[capitalize,noabbrev]{cleveref}
\crefformat{footnote}{#2\footnotemark[#1]#3}

\RequirePackage{xstring}
\RequirePackage{xparse}
\RequirePackage[]{acro}
\NewDocumentCommand\acrodef{mO{#1}mG{}}{\DeclareAcronym{#1}{short={#2}, long={#3}, #4}}

\DeclareAcronym{SAA}{short=SAA, long=sample average approximation}

\DeclareAcronym{DM-RS}{short=DM-RS, long=demodulation reference signals}

 \DeclareAcronym{APES}{short=APES, long=amplitude and phase estimation}
 
\DeclareAcronym{MF}{short=MF, long=matched filter}
\DeclareAcronym{QoS}{short=QoS, long=quality of service}
\DeclareAcronym{KF}{short=KF, long=Kalman filter}
\DeclareAcronym{HDA}{short=HDA, long= hybrid digital-analog}
\DeclareAcronym{BT}{short=BT, long= beam trackin}
\DeclareAcronym{RSU}{short=RSU, long=roadside unit}
\DeclareAcronym{mMIMO}{short=mMIMO, long=massive multi-input multi-output}

\DeclareAcronym{(V2I}{short=(V2I, long= Vehicle-to-Infrastructure}
 
 \DeclareAcronym{SCA}{short=SCA, long= successive convex approximation}

\DeclareAcronym{THz}{short=THz, long= terahertz}

 \DeclareAcronym{SE}{short=SE, long= spectral efficiency}  

  \DeclareAcronym{BW}{short=SE, long= bandwidth}  
  
\DeclareAcronym{V2X}{short=V2X, long= vehicle-to-everything}
    
\DeclareAcronym{EKF}{short=EKF, long=extended Kalman filter}

\DeclareAcronym{MMSE}{short=MMSE, long=minimum mean square error}
\DeclareAcronym{OFDM}{short=OFDM, long=orthogonal frequency division multiplexing}
\DeclareAcronym{NOMA}{short=NOMA, long=non-orthogonal multiple access}
\DeclareAcronym{SIC}{short=SIC, long=successive interference cancellation}
\DeclareAcronym{URLLC}{short=URLLC, long=ultra-reliable low-latency communications}

\DeclareAcronym{MPC}{short=MPC, long=multipath component}
\DeclareAcronym{6G}{short=6G, long=sixth generation}
\DeclareAcronym{5G}{short=5G, long=fifth generation}
\DeclareAcronym{CSI}{short=CSI, long=channel state information}
\DeclareAcronym{GLRT}{short=GLRT, long=generalized likelihood ratio test}
\DeclareAcronym{iid}{short=i.i.d., long=independent identically distributed}
\DeclareAcronym{mmWave}{short=mmWave, long=millimeter wave}
\DeclareAcronym{MUSIC}{short=MUSIC, long=multiple signal classification}
\DeclareAcronym{CRB}{short=CRB, long= Cramér-Rao bound}
\DeclareAcronym{MSE}{short=MSE, long=mean squared error}
 
 \DeclareAcronym{MCD}{short=MCD, long=multipath-component-distance}
\DeclareAcronym{V2V}{short=V2V, long=vehicle-to-vehicle}

\DeclareAcronym{RCS}{short=RCS, long=radar cross-section}
\DeclareAcronym{SI}{short=SI, long=self-interference}
\DeclareAcronym{SNR}{short=SNR, long=signal-to-noise ratio}
\DeclareAcronym{SVD}{short=SVD, long=singular value decomposition}
\DeclareAcronym{ULA}{short=ULA, long=uniform linear array}
\DeclareAcronym{MINLP}{short=MINLP, long=mixed-integer linear program}
\DeclareAcronym{SDR}{short=SDR, long=Semidefinite relaxation}

\DeclareAcronym{HAD}{short=HAD, long=hybrid analog-digital}

\DeclareAcronym{ISAC}{short=ISAC, long=integrated sensing and communications}
\DeclareAcronym{BS}{short=BS, long=base station}
\DeclareAcronym{SINR}{short=SINR, long=signal-to-interference-plus-noise ratio}
\DeclareAcronym{AWGN}{short=AWGN, long=additive white Gaussian noise}
\DeclareAcronym{FD}{short=FD, long=full-duplex}

\DeclareAcronym{DFRC}{short=DFRC, long=dual-functional radar and communication}
\DeclareAcronym{AoA}{short=AoA, long=angle of arrival}
\DeclareAcronym{AoD}{short=AoD, long=angle of departure}
\DeclareAcronym{UE}{short=UE, long=user equipment}
\DeclareAcronym{MIMO}{short=MIMO, long=multiple-input multiple-output}
\DeclareAcronym{CDF}{short=CDF, long=cumulative distribution function}
\DeclareAcronym{NLoS}{short=NLoS, long=non-line-of-sight}
\DeclareAcronym{LoS}{short=LoS, long=line-of-sight}

\DeclareAcronym{SC}{short=S\&C, long=sensing and communication}

\DeclareAcronym{SOC}{short=SOC, long=second-order cone}

\DeclareAcronym{KRST}{short=KRST, long=Khatri-Rao space-time}

\DeclareAcronym{V2I}{short=V2I, long=Vehicle-to-Infrastructure}

\DeclareAcronym{NR}{short=NR, long=new radio}
\DeclareAcronym{RMSE}{short=RMSE, long=root mean square error}

\DeclareAcronym{RCRB}{short=RCRB, long=root Cramér-Rao bound}

\DeclareAcronym{PSD}{short=PSD, long=positive semidefinite}

\usepackage{bm}
\usepackage{algorithm}
\usepackage{algpseudocode}
\usepackage{comment}
\usepackage{transparent}
\usepackage{todonotes}

\NewDocumentCommand\IEEE{ s m d[] }{%
	\IfBooleanTF{#1}{}{IEEE\,}% suppress IEEE when using starred form
	\nolinebreak[2]% this is a somewhat bad place for a line break
	#2%
	\IfNoValueTF{#3}{%
		}{%
		\StrGobbleLeft{#3}{1}[\sommerIEEEFirstLetter]%
		\IfEq{\sommerIEEEFirstLetter}{}{%
			#3% just one letter, do not allow line break
			}{%
			\nolinebreak[3]% multiple letters, this is just a very bad place for a line break
			\StrLeft{#3}{1}%
			\sommerIEEELettersSlashed{\sommerIEEEFirstLetter}%
		}%
	}%
}
\newcommand{\sommerIEEELettersSlashed}[1]{%
	/% separate letters by slashes
	\StrLeft{#1}{1}%
	\StrGobbleLeft{#1}{1}[\sommerIEEESubsequentLetter]%
	\IfEq{\sommerIEEESubsequentLetter}{}{%
		}{%
		\sommerIEEELettersSlashed{\sommerIEEESubsequentLetter}% recurse
	}%
}

\begin{document}

\title{Toward Reliable and Accurate Predictive ISAC in Mobile mmWave Networks}

\author{  Atefeh Rezaei, \textit{Member, IEEE}, Vahid Jamali, \textit{Senior Member, IEEE}, and Falko Dressler, \textit{Fellow, IEEE}\thanks{ A. Rezaei and F. Dressler are with the School of Electrical Engineering and
Computer Science, Technische Universität Berlin, Germany (e-mail: rezaei@ccs-labs.org, dressler@ccs-labs.org). }
\thanks{ V. Jamali is with the Resilient Communication
Systems Laboratory,  Technische Universität Darmstadt,  64283 Darmstadt, Germany (e-mail: vahid.jamali@tu-darmstadt.de).}
\thanks{This work was supported by the Federal Ministry of Research, Technology,
and Space (BMFTR, Germany) within the project xG-RIC under grant
16KIS2429K as well as by the German Research Foundation (DFG) within
the project RADCOM-HETNET under grant DR 639/18-4.}} 

\maketitle

\begin{abstract}\nohyphens{%
\Ac{ISAC} systems offer a promising framework for beam tracking, which enhances channel awareness and improves communication reliability in mobile \ac{mmWave} networks. 
Motivated by this, we propose a low-complexity, model-driven \ac{CSI} framework with integrated channel prediction capability in \ac{ISAC}-enabled systems.
To reduce the overhead of \ac{CSI} acquisition and uplink feedback, the proposed framework leverages predicted target sensing parameters to construct partial \ac{CSI} from both \ac{LoS} and \textit{detected} \ac{NLoS} components. 
Exploiting this partial CSI, we formulate a beamforming optimization problem that minimizes the worst targets' \ac{CRB} under per-user communication constraints.
This problem is non-convex and highly non-linear; therefore, we develop an efficient suboptimal solution using \ac{SCA} and Schur-complement-based decomposition techniques, and we evaluate its complexity in comparison with other approaches.
Moreover, since obtaining the absolute phase of the channel paths may require additional pilot resources,  we formulate a robust beamforming design that does not assume this knowledge, and propose a solution based on sample averaging.
Numerical results demonstrate that the \ac{ISAC}-based robust channel prediction and localization framework significantly improves prediction accuracy and link reliability compared with conventional schemes, such as communication-based localization and tracking and separated sensing and communication.
These results highlight the potential of \ac{ISAC} for reliable, adaptive, and high-frequency, blockage-prone wireless networks.
Furthermore, the performance of the proposed framework is evaluated under severe blockage scenarios, which further demonstrates enhanced network reliability.
}\end{abstract}


\acresetall%
\IEEEpeerreviewmaketitle%

\section{Introduction}

\Ac{6G} mobile networks are anticipated to support a wide range of high-precision sensing services by leveraging technologies such as \ac{ISAC}. 
In particular, \ac{ISAC}-enabled beam tracking can simultaneously support accurate localization and tracking, along with high-speed data exchange, which allows real-time environmental monitoring for applications such as intersection management and collision avoidance in high-speed vehicular scenarios \cite{liu2020joint}.
%Moreover, the wide bandwidth and high spatial resolution of \ac{mmWave} and massive \ac{MIMO} technologies further enable \ac{DFRC} systems to achieve precise positioning and high-rate communication simultaneously~\cite{liu2020joint-radar}.
Accurate \ac{CSI} is essential in modern wireless systems, particularly in \ac{mmWave} and \ac{mMIMO} systems.
Additionally, the large antenna arrays and the sparse propagation characteristics of \ac{mmWave} channels make conventional estimation schemes or \ac{5G}-\ac{NR} approaches prohibitively costly in terms of pilot overhead.
In this regard, a common approach is to combine pilot signals with codebook-based beam training. 
While this method balances estimation accuracy and overhead, it leads to an inherent tradeoff: richer codebooks improve performance but increase training cost \cite{alkhateeb2014channel,hur2013millimeter,xiao2016hierarchical}.

Unlike traditional feedback-based beam tracking methods, \ac{DFRC} signals significantly reduce signaling overhead and improve the accuracy of angle estimation \cite{yuan2021bayesian,liu2020radarassisted}.
Instead of relying on exhaustive beam probing, \ac{ISAC} enables location- and environment-aware \ac{CSI} estimation, which reduces overhead and enhances robustness in mobile scenarios.
However, its practical implementation requires advanced waveform and hardware design, as well as reliable sensing-to-\ac{CSI} mapping mechanisms.
Recent studies have demonstrated that \ac{ISAC} can extract spatial features such as \ac{AoA}, \ac{AoD}, and Doppler, which can be mapped to communication-relevant \ac{CSI} \cite{liu2020joint,du2023integrated}.

% \textbf{Contributions:}
Compared with conventional pilot-based methods, we deploy predictive \ac{ISAC} to reduce overhead and eliminate the need for uplink feedback.
However, acquiring accurate channel information under \ac{NLoS} conditions is particularly challenging. This motivates the development of channel models that capture such dynamics while retaining the low-complexity and low-overhead advantages of \ac{ISAC}-based tracking. Existing literature on sensing-assisted tracking often overlooks accurate channel modeling under medium to high user mobility and fails to address scenarios that require both precise channel representation and efficient optimization.
To address this, we propose a model-driven \ac{CSI} recovery framework under \ac{NLoS} conditions. 
In this framework, \ac{ISAC}-based sensing extracts geometric information, which is then incorporated into the channel via geometric reconstruction, significantly reducing channel estimation and prediction overhead. Our model accounts for \ac{LoS} paths, fast-changing \ac{NLoS} components detected through \ac{ISAC}, and other non-detected scatterers. Inter-vehicle reflections are explicitly modeled as mobile scatterers that capture dynamic propagation effects.
By combining \ac{ISAC}-derived sensing outputs with an \ac{EKF}, the proposed framework enables high-accuracy, low-overhead \ac{CSI} prediction. With a focus on \ac{mmWave} systems, we leverage their sparse multipath nature, allowing geometry-based channel models to efficiently capture propagation behavior.  The key novelty of this work is a geometry-aware channel prediction framework that reconstructs both the LoS component and dynamically varying NLoS components associated with mobile scatterers, and exploits the predicted channel for robust next-slot beamforming.

To mitigate practical impairments such as carrier frequency offset, multipath fading, Doppler shift, and time synchronization errors, which induce phase rotations in received signals, we adopt a non-coherent beamforming design. This approach eliminates pilot overhead and enhances robustness in mobile networks, even when uplink feedback is absent or impaired. Building on this, we develop a robust beamforming design that ensures reliable performance despite unknown relative phase variations. Furthermore, we propose two optimization schemes for the coherent and non-coherent cases, designed to improve estimation and tracking accuracy by minimizing the worst-user \ac{CRB} while maintaining high-quality data communication.

Our key contributions can be summarized as follows:
\begin{itemize}
\item We develop a low-overhead \ac{CSI} framework, which exploits the predicted sensing parameters of the targets to construct partial \ac{CSI} from both \ac{LoS} and \textit{detected} \ac{NLoS} components.
\item We propose a dynamic beamforming design that minimizes the worst-user \ac{CRB} via epigraph- and Schur-based reformulation, using surrogate functions to convexify \ac{CRB} and rate constraints.
\item As obtaining the absolute phase of the channel paths requires extra pilot resources, signaling, and feedback, we formulate a robust beamforming design that does not assume this knowledge, and propose a solution based on sample averaging.
\item A comprehensive performance evaluation demonstrates the advantages of \ac{ISAC}-based localization and tracking, highlights the importance of \ac{NLoS} links when \ac{LoS} paths are blocked, and quantifies the gains achieved through robust optimization under mobile, non-coherent conditions.
\end{itemize}

The rest of the paper is organized as follows.
Section \ref{relatedwork} provides an overview of related works. Section \ref{systemmodel} presents the proposed system model, including the radar and communication receiver models. In Section \ref{framework}, we describe the proposed predictive \ac{ISAC} framework, which covers the tracking and channel modeling modules, as well as the formulation of our optimization problem and its solution methodology for both coherent and non-coherent cases. We also analyze the overall algorithm, including convergence and computational complexity. Numerical results obtained through simulations are presented in Section \ref{simulations} to evaluate the performance of the proposed framework and solution. Finally, Section \ref{conclusion} concludes the paper.
  
\textit{Notations}:
Throughout this paper, scalars are denoted by lowercase letters (e.g., $a,b$), vectors by bold lowercase letters (e.g., $\mathbf{h}, \mathbf{w}$), and matrices by bold uppercase letters (e.g., $\mathbf{H}, \mathbf{W}$). The superscripts $(\cdot)^T$ and $(\cdot)^H$ denote the transpose and Hermitian (conjugate transpose) of a matrix or vector, respectively. The derivative of a vector $\mathbf{a}$ with respect to time is denoted by $\dot{\mathbf{a}}$. The absolute value and Euclidean norm of a scalar or vector are denoted by $|\cdot|$ and $\|\cdot\|$, respectively. The symbol $\mathbf{a} \circ \mathbf{b}$ denotes the Hadamard (element-wise) product between vectors or matrices, and $\mathcal{R}(\cdot)$ denotes the real part of the argument. The number of linearly independent rows of a matrix is denoted by $\text{rank}(\cdot)$, and $\mathbf{A} \succeq \mathbf{0}$ indicates that $\mathbf{A}$ is \ac{PSD}.

\section{Related Work}\label{relatedwork}

In the context of \ac{ISAC}-based tracking applications, Liu et al. \cite{liu2020joint} were among the first to propose an innovative transceiver design and frame structure for a \ac{DFRC}-\ac{BS} operating in the \ac{mmWave} spectrum.
The proposed system enables the \ac{BS} to simultaneously serve a multi-antenna \ac{UE} and detect targets. 
On the other side, Yuan et al. \cite{yuan2021bayesian} introduced a method in which a \ac{RSU} infers vehicle motion parameters directly from \ac{DFRC} signal echoes, which reduces signaling requirements and improves tracking precision.
Liu et al. \cite{liu2020radarassisted} proposed a radar-assisted predictive beamforming design for \ac{V2I} communication based on \ac{DFRC}, which applies an \ac{EKF} at the \ac{UE} to track and predict vehicle kinematics, while jointly optimizing transmission power for data rate and posterior \ac{CRB}-based parameter estimation.

Li et al. \cite{li2025ekfbased} developed a beamforming design for conventional joint beam tracking and communication systems, shifting the focus from the power allocation approach previously investigated in \cite{liu2020radarassisted}.
Similarly, Du et al. \cite{du2023integrated} investigate sensing-assisted beamforming for \ac{V2I} communication, where a \ac{RSU} with a \ac{mMIMO} array at \ac{mmWave} frequencies tracks an extended target.
Their scheme dynamically adjusts the beamwidth for full vehicle coverage and uses an \ac{EKF} to track the vehicle’s position via resolved scatterers, while optimizing the time-splitting factor to define wide and narrow beamwidth levels.
Liu et al. \cite{liu2025multipath} employ \ac{KRST} code to exploit multipath components in a single \ac{ISAC} system, which enables high-accuracy multi-target sensing while supporting multi-user communication.
The \ac{KRST} code allows flexible trade-offs between communication and sensing performance.

Fernando et al. \cite{fernando2025sensing}  propose estimation and tracking algorithms for analog, digital, and \ac{HDA} antenna architectures, focusing on received beamforming design.
For digital and hybrid beamforming, predictions are derived from a Markov chain model used for analog beamforming. 
Zhao et al. \cite{zhao2023sensing-assisted} use \ac{ISAC}-based beam tracking to monitor the user’s position and employ a binary hypothesis test to detect obstacles between the \ac{BS} and the vehicle. When an obstacle is detected, the system dynamically switches from \ac{mmWave} to Sub-6 GHz communication to maintain reliable connectivity.
Rezaei et al. \cite{rezaei2026mobility-aware}  investigated an interference-coupled ISAC framework for multi-vehicle mmWave tracking with a focus on energy-efficient power allocation under explicit inter-vehicle interference modeling. In contrast, the present work addresses CRB-optimal robust beamforming under LoS/NLoS channel uncertainty and unknown absolute phase.
\begin{table*}[t]
\centering
\caption{Comparison with related work}
\label{tab:comparison related work}
\begin{tabular}{p{1.5cm}p{1.2cm}p{1.4cm}p{1.3cm}p{1.3cm}p{1.7cm}p{1.6cm}}
    \toprule
    \textbf{Reference} & \textbf{Interf.-aware} & \textbf{Coherent Beamforming} & \textbf{Non-coherent Beamforming} & \textbf{Blockage detect.} & \textbf{Accurate CSI-driven model: inc. NLoS} & \textbf{Type} \\
    \midrule
    \cite{liu2020radarassisted} & - & Yes (power) & - & - & - & Sens. centric \\
    \cite{du2023integrated} & - & Yes & - & - & - & Sens. centric \\
    \cite{fernando2025sensing} & Yes & Yes & - & - & - & Com. centric \\
    \cite{zhao2023sensing-assisted} & - & Yes & - & Yes & - & Com. centric \\
    \cite{li2025ekfbased} & Yes & Yes & - & - & - & Sens. centric \\
    \cite{liu2025multipath} & Yes & Yes & - & - & - & Sens. centric \\
    \cite{rezaei2026mobility-aware} & Yes & Yes (power) & - & - & - & Com. centric \\
    \midrule
    \textbf{Ours} & Yes & Yes & Yes & Yes & Yes & Sens. centric \\
    \bottomrule
\end{tabular}
\end{table*}

Unlike conventional and \ac{5G}-\ac{NR}-based estimation and tracking approaches, this work focuses on low-overhead channel prediction by leveraging ISAC-enabled predictive beamforming. It is worth noting that the aforementioned \ac{DFRC}-based beam-tracking studies predominantly assume \ac{LoS} channels, without incorporating accurate channel models that capture the contributions of non-line-of-sight components.
Moreover, prior works such as \cite{yuan2021bayesian,du2023integrated,fernando2025sensing,zhao2023sensing-assisted} primarily focus on optimizing communication data rates, while paying limited attention to whether the reflected signal strength is sufficient to ensure reliable and high-quality sensing, thereby leaving a gap for the development of novel optimization strategies. 

 Table \ref{tab:comparison related work} highlights the research gap
addressed in this paper from the system-model perspective. In
particular, unlike existing approaches that primarily focus on
LoS-dominated or static multipath components, the proposed framework
enables geometry-aware channel prediction by explicitly modeling
dynamically varying NLoS components arising from mobile scatterers.
Specifically, while \cite{zhao2023sensing-assisted} detects LoS blockage and falls back to
Sub-6 GHz operation to maintain connectivity, our framework exploits
the blockage event differently: the detected mobile scatterers are
used to reconstruct the NLoS components of the channel, allowing the
blocked user to continue being served over mmWave links rather than
requiring a band switch. Similarly, \cite{liu2020radarassisted} and \cite{du2023integrated} exploit sensing for
predictive beam tracking but assume LoS-dominated channels and
coherent phase knowledge. Among the considered works, the proposed
framework is the only one that combines an accurate CSI-driven model
including NLoS components, non-coherent beamforming design addressing
absolute phase ambiguity, blockage detection, and explicit
interference-aware predictive beamforming within a single ISAC
tracking framework.

The proposed problem formulation differs from existing
predictive ISAC designs: whereas \cite{liu2020radarassisted} and \cite{rezaei2026mobility-aware} optimize transmit power
allocation and \cite{du2023integrated} optimizes a time-splitting factor, we directly
design the dual-functional beamforming vectors to minimize the
worst-user CRB subject to per-user rate constraints, thereby
guaranteeing a sensing accuracy bound for every tracked target rather
than optimizing an aggregate metric. To solve the resulting non-convex
min-max problem, we develop an epigraph- and Schur-complement-based
reformulation combined with SCA, and additionally formulate a robust
non-coherent variant that does not require absolute phase knowledge.

\section{System Model}\label{systemmodel}

We assume a \ac{DFRC} system \cite{chen2022generalized,liu2020transmit} that employs a single transceiver and waveform to simultaneously perform wireless communication and radar sensing, thereby extracting motion and position information from reflected signals. In this system, the radar transmits probing signals and receives reflections from various objects in the environment.
Specifically, we consider vehicles moving along a street network at different positions and speeds, which reflect the \ac{DFRC} signals back to the \ac{DFRC}-\ac{BS} and, potentially, to other users during signal reception.  
Consequently, each vehicle acts as a potential mobile scatterer for nearby users, influencing the received signals in the networked vehicular environment.

\begin{figure}
    \centering
    \def\svgwidth{\columnwidth}
    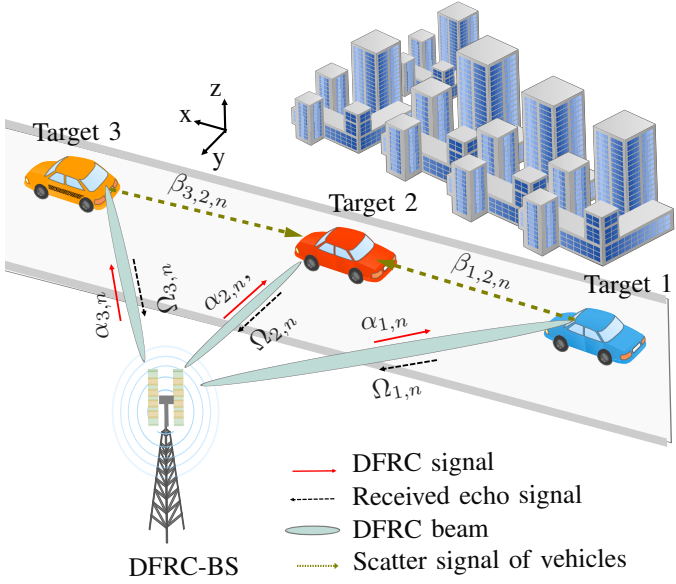
    \caption{Joint communication and sensing in a mobile network.}%
    \label{fig:sys}
\end{figure}

The \ac{MIMO} \ac{DFRC}-\ac{BS} is equipped with two \acp{ULA}.
The system comprises $N_t$ transmit antennas and $N_r$ receiving elements, which provides downlink communication to $K$ potential mobile users while simultaneously sensing the environment to track them, as depicted in \Cref{fig:sys}.
Here, $\Omega_{k,n}$ corresponds to the sensing round-trip channel, while $\alpha_{k,n}$ and $\beta_{i,k,n}$ denote the path loss between the \ac{BS} and a vehicle, and between two different vehicles, respectively.
The information collected on these targets improves communication performance.

We consider receive beamforming at the \ac{BS} based on the predicted user angle, obtained from the tracking module. By steering the receive beam toward the predicted user direction, the beamformer naturally places low gain in other directions, thereby attenuating clutter originating from undesired directions \cite{rosadosanz2022adaptive,rojhani2024comprehensive}. In addition to receive beamforming, high-resolution angle estimation techniques such as \ac{MUSIC} can further discriminate between user signals and clutter.
\Ac{MUSIC} exploits the spatial covariance of the received signal to estimate the angles of arrival of multiple paths with high precision. By combining predicted user angles from the tracking module with \ac{MUSIC}-based angle estimates, the \ac{BS} can both reinforce the desired user signals and suppress clutter from environmental reflections. This integration of tracking and \ac{MUSIC} allows the system to distinguish closely spaced paths, enhancing the reliability of the channel estimation in dense or cluttered vehicular environments.
In addition to this directional filtering, we assume that, due to the large \ac{BS}-\ac{UE} distance, multi-bound reflections are largely mitigated.
For other types of mobile vehicles as scatterers without communication requirements, we treat them as users with a zero data-rate requirement, while still sensing them to obtain their geometric information, which can aid in accurate channel modeling and accident avoidance.

\subsection{Radar Model}\label{radar_m}

We consider a total tracking frame duration $T$, which is divided into 
$N$ discrete time slots, each with the duration of $\delta_t$.
The total transmitted signal from the \ac{DFRC}-\ac{BS} at time instant $t\in[0,\delta_t)$ of the $n$-th time slot according to all users is expressed as:
\begin{align}\label{transmit_signal}
\mathbf{s}_n(t)&= \sum_{k\in \mathcal{K}} \mathbf{w}_{k,n}s_{k,n}(t),
\end{align}
where $s_{k,n}(t)$ is the transmitted signal to individual user $k$ in time slot $n$.
Moreover, we assume the dual-functional beamforming vector for the corresponding signal $s_{k,n}(t)$ as $ \mathbf{w}_{k,n}\in \mathrm{C}^{N_t\times 1}$.
We define $\mathbf{a}_t(\mathbf{\theta})$ and  $\mathbf{a}_r(\mathbf{\theta})$ as the steering vectors at the transmit and receive radar antennas, respectively.
The angle of departure, $\mathbf{\theta}$, corresponds to the locations of tracked users, and $\lambda$ is the carrier wavelength. 
$\hat{d}$ represents the spacing between two adjacent antennas. Consequently, the steering vectors are defined as:
\begin{align} \label{equ:steering}
	\hspace{-0.25mm}\mathbf{a}_{t}(\mathbf{\theta})\hspace{-0.25mm}=& \hspace{-0.25mm}\frac{1}{\sqrt{N_t}}\big[1,e^{-j 2\pi \frac{\hat{d}}{\lambda} \cos (\theta)},..., e^{-j 2\pi \frac{\hat{d}}{\lambda} (N_{t}-1) \cos(\theta)}\big]^T
\end{align} 
\begin{align} \label{equ:steeringr}
	\hspace{-0.25mm}\mathbf{a}_{r}(\mathbf{\theta})\hspace{-0.25mm}=& \hspace{-0.25mm}\frac{1}{\sqrt{N_r}}\big[1,e^{-j 2\pi \frac{\hat{d}}{\lambda} \cos (\theta)},..., e^{-j 2\pi \frac{\hat{d}}{\lambda} (N_{r}-1) \cos(\theta)}\big]^T
\end{align} 

%\subsubsection{Received Echo Signal}
Let $\mathbf{H}_{k,n}\hspace{-1mm}=\Omega_{k,n} \mathbf{a}_{r}{(\theta_{k,n})}\mathbf{a}_{t}^{H}{(\theta_{k,n})}$ represent the round-trip channel matrix of the $k$-th target, which is assumed to be fixed within the entire time slot.
Here, $\Omega_{k,n}$ denotes the complex-valued channel coefficient, which depends on the target's \ac{RCS}, denoted by $\nu_{\text{RCS}}$, and the signal propagation distance where $\Omega_{k,n}=\nu_{\text{RCS}}(2d^{-1}_{k,n})$. 
  The RCS of the vehicle
is assumed to be constant, corresponds to a Swerling I target \cite{liu2020joint}.
Considering \eqref{transmit_signal}, the received reflected echoes at time slot $n$ of the radar can be expressed as:
\begin{equation}\label{re_sens1}
\mathbf{r}_{n}(t)= \sqrt{N_tN_r}\sum_{k \in \mathcal{K}}e^{j2\pi\mu_{k,n}t}\mathbf{H}_{k,n}\mathbf{s}_n(t-\tau_{k,n})+\mathbf{z}(t),
\end{equation}
where $ \sqrt{N_tN_r} $ is the antenna gain and $\mathbf{z}\sim\mathcal{C}\mathcal{N}(\mathbf{0},\sigma^{2}\mathbf{I}_{N_r})$ represents the received \ac{AWGN} at the \ac{DFRC}-\ac{BS} in each individual time slot.
First, receive beamforming is applied at the radar side to suppress clutter and enhance the target reflection.
Accordingly, we define the receive beamforming vector as $\mathbf{b}_{k,n}=\mathbf{a}_{t}^{H}{(\theta_{k,n|n-1})}$, which is designed based on the predicted angle from the previous time slot.
This approach provides better alignment with the expected direction of arrival, accounting for the user’s mobility.
Then, the received signal at \ac{DFRC}-\ac{BS} in time slot 
$n$ according to target $k$ is: 
\begin{align}\label{re_sens2}
&
{r}_{k,n}(t)= \mathbf{b}^H_{k,n} \mathbf{r}_{n}(t)\nonumber \\ &=\sqrt{N_tN_r} e^{j2\pi\mu_{k,n}t}\mathbf{b}^H_{k,n}\mathbf{H}_{k,n}\mathbf{s}_n(t-\tau_{k,n})+\mathbf{b}^H_{k,n}\mathbf{z}(t),\forall k, n,
\end{align}
where $\mathbf{b}^H_{k,n}=\mathbf{a}_r(\theta_{k,n|n-1})$ is the steering vector according to the predicted angle of the vehicle $k$ based on the time slot $n-1$. 
%%%%%%%%%%%%%%%%%%%%%%%%%%%%%%%%%%%%%%%%%%%%%%
\subsection{Communication Receiver Model}\label{rate_com}
It is worth noting that the described \ac{DFRC} system with \ac{ISAC} signal simultaneously carries communication data over the entire transmission block.
This design not only allocates more time for the sensing operation but also ensures uninterrupted data transmission within each time slot.
Thus, given the transmitted signal described in \eqref{transmit_signal}, at the 
$n$-th epoch, the $k$-th vehicle captures the signal emitted by the \ac{BS} as follows:
\begin{equation}\label{re_user}
{y}_{k,n}(t)= \sqrt{N_t}\sum_{i \in \mathcal{K}}\mathbf{h}^H_{k,n}\mathbf{w}_{i,n}s_{i,n}(t)+n_{k,n},
\end{equation}
where $\sqrt{N_t}$ depicts the array gain factor and $n_{k,n}\sim\mathcal{C}\mathcal{N}(0,\sigma_{k,n}^2)$ represents the \ac{AWGN} where $\mathcal{C}\mathcal{N}$ denotes a circularly symmetric complex Gaussian distribution with zero mean and variance $\sigma_{k,n}^2$.
We note that the received signal at each user consists of a direct-path component as well as reflections from surrounding objects and other vehicles.
Therefore, we propose the channel model for the  $k^{\text{th}}$ user, $\mathbf{h}_{k,n}$, to include both the \ac{LoS} link and the \ac{NLoS} components, which are estimated based on the angle, Doppler shift, delay, and location of the $K$ tracked users.
We describe $\mathbf{h}_{k,n}$ later in the channel modeling section in more detail.
Considering the received signal at the user's side, the received SINR for each user is described as:
\begin{equation}\label{C_SINR}
\Gamma_{k,n}=\frac{\left |\sqrt{N_t}\mathbf{h}_{k,n}^H\mathbf{w}_{k,n}\right |^2}{\hspace*{-3mm}\underset{i\in\mathcal{K}\setminus \left \{ k \right \}}{\sum }\left |\sqrt{N_t}\mathbf{h}_{k,n}^H\mathbf{w}_{i,n}\right |^2 +  \sigma^2_{k,n}},~~\forall k, n.  
\end{equation}
Based on \eqref{C_SINR}, the achievable data rate at the $k$-th vehicle is described as $R_{k,n}=\log(1+\Gamma_{k,n})$.

\section{Proposed Predictive \ac{ISAC} Framework}\label{framework}

\begin{figure*}
    \centering
    \def\svgwidth{0.75\textwidth}
    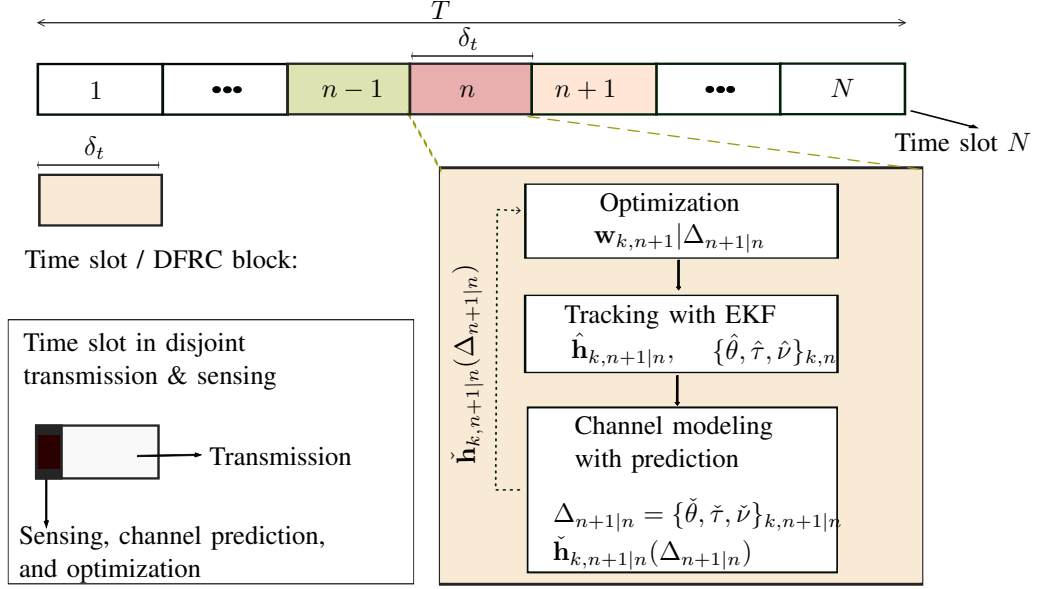
    \caption{ Proposed tracking frame structure, where each time slot includes a DFRC block for joint data transmission and echo signal processing, enabling channel prediction and predictive beamforming for the next time slot.}%
    \label{frame}
\end{figure*}

 This section provides the proposed frame structure for \ac{ISAC}-assisted networks, illustrated in \Cref{frame}. 
We assume a total tracking time/time frame $T$  partitioned into $N$ time slots defined by the \ac{DFRC}-\ac{BS} scheduler as the central unit\footnote{All vehicles are assumed to be synchronized to the BS clock via standard network synchronization mechanisms (e.g., cellular signaling), ensuring consistent time alignment for sensing, communication, and state tracking operations \cite{3gpp-38211,lin2025integrated}.}. We assume a time slot duration of $\delta_t = 0.02$ s, consistent with prior work on ISAC-based prediction and beam management in \ac{mmWave} networks \cite{liu2020radarassisted,du2023integrated}.
 Compared with separate sensing and communication, we consider a DFRC  block, where the same waveform is used for both sensing and communication within each time slot. During a time slot, we transmit data to the UEs while simultaneously performing signal processing on the received echo signals to extract UE features. Using this sensing information, we then predict the characteristics for the next time slot, including the channel state and predictive beamforming parameters.

We highlight that in \Cref{framework}, during each time slot, the \ac{BS} simultaneously transmits data to vehicles using the optimized predicted beamforming vector and collects the reflected echoes over the entire slot. The received signals are processed via \ac{MF} and \ac{MUSIC} to estimate the channel parameters. These estimates are then refined by an \ac{EKF} to predict the channel state in the next slot. Based on the predicted channel, the beamforming optimization is performed to generate the transmit beamforming for the subsequent time slot, thereby closing the ISAC loop.

Based on this framework (\Cref{frame}), we divide the operations in each time slot into three sequential modules, where each module depends on the outputs of the previous one, as described in the following.

\subsection{Tracking Module}

 The tracking module is a functional block within the system responsible for tracking and consists of several sequential tasks, as follows: 
(a) At the $n$-th time slot, the \ac{DFRC}-\ac{BS} emits $K$ transmit beams towards targets, utilizing  predictions of $ {\theta}_{k,n|n-1}$, $ {d}_{k,n|n-1}$, $ {\nu}_{k,n|n-1}$, $\Omega_{k,n|n-1}$, which are based on the state evaluation models.
(b) The \ac{DFRC}-\ac{BS}  receives the signal reflected by the targets described in \eqref{re_sens2}.  Considering the received signal at the \ac{BS}, we apply the \ac{MUSIC} algorithm \cite{schmidt1986multiple}, which offers high-resolution angle estimation, crucial for accurately tracking multiple users in the presence of clutter.  
Accurate angle estimation is critical because it allows the beamforming and tracking algorithms to correctly associate reflections with the corresponding vehicles. Afterward, the \ac{MF} algorithm \cite{li2021joint} is applied to estimate the delay and Doppler shifts of the targets. This separation of tasks is used because \ac{MUSIC} provides high angular resolution, whereas the \ac{MF} algorithm efficiently extracts delay or Doppler information once the angles are known. 
 (c) Considering a state prediction and tracking approach, we employ an \ac{EKF} to track and predict the kinematic parameters of each vehicle, including the angle, distance, and velocity under the vehicle kinematic model.

  The considered \ac{EKF} based on \cite{liu2020radarassisted} is employed for beam prediction and tracking. It locally linearizes the nonlinear state-transition and measurement models to manage the nonlinear relationships between the vehicle’s kinematic states—such as \ac{AoA}, propagation delay, and Doppler shift—and the observed measurements. 
The adopted \ac{EKF}, applied in the context of beam tracking for each mobile vehicle, refines the estimation accuracy of angle, distance, Doppler, and channel coefficient (${\hat{\theta}_{k,n}, \hat{d}_{k,n}, \hat{\nu}_{k,n}, \hat{\Omega}_{k,n}}$) at the $n$-th time slot.  
 These refined state parameters are then used as inputs for the beam/kinematic state predictor at the $(n+1)$-th epoch at the \ac{DFRC}-\ac{BS}. Specifically, the predictor leverages the estimated angle, delay, and Doppler from the \ac{EKF} to forecast the next positions and channel parameters of each mobile vehicle, enabling more accurate beamforming. We remark that the \ac{DFRC}-BS  is responsible for target tracking, which is crucial for accurately modeling the channel.

\subsection{Channel Modeling Module}

 Given the steering
vector, the channel can be described by a multi-path
model as \cite{tse2005fundamentals,heath2016overview}
\begin{align}\label{channel_full}
\mathbf{h}_{k,n}(t) &= \sum_{\ell=1}^{G} 
g_{k,n,\ell} \, e^{j2\pi \nu_{k,n,\ell} t} \, \mathbf{a}_{(\theta_{\ell,n})},
\end{align} 
where $g_{k,n,\ell}$, $\nu_{k,n,\ell}$, and $\theta_{\ell,n}$ denote the complex channel gain, Doppler shift, and steering angle associated with the $\ell$-th path, respectively, and $G$ is the total number of paths between the BS and the $k$-th user at $n$-th epoch and time $t$. 
Various channel estimation methods have been proposed in the literature.
Conventional channel estimation methods include least-squares or compressive sensing techniques  \cite{tse2005fundamentals} followed by an uplink feedback to the BS.
For mmWave systems with large arrays, the channel dimension increases, and sparsity features arise (i.e., only a few dominant paths exist leading to small $G$). 
In such cases, compressive sensing methods or channel parameter estimation followed by channel reconstruction are typically employed \cite{el-ayach2014spatially,gao2016channel}.

 In 5G NR, beam management requires significant pilot overhead due to channel state information reference signals (CSI-RS) transmission and uplink feedback. ISAC-based beam tracking offers a means to reduce this overhead. Conventional schemes require \ac{RSU}-side pilot transmission before each downlink data block, followed by vehicle-side channel estimation and uplink feedback. Sensing-assisted approaches alleviate this burden. However, \ac{DM-RS} remains essential for channel estimation and coherent demodulation  \cite{du2023integrated}.
 By exploiting geometry-aware \ac{ISAC} information, the proposed framework avoids the feedback requirement from the user to the \ac{DFRC}-\ac{BS} for exhaustive estimation of all multipath parameters. Hence, the proposed framework reconstructs only the detectable channel components and predicts the channel for the next time slot, making it particularly well-suited for mobile \ac{mmWave} vehicular networks and enabling predictive beamforming based on a nearly accurate channel without requiring any user feedback.

The key idea of the proposed channel modeling module is that, in the considered multi-user ISAC system, part of the channel components in \eqref{channel_full} can be inferred from target location information.
This includes the LoS component and the NLoS components produced by targets that act as reflectors for other nodes.
In the following, we formalize the known and unknown components of the channel. 
 We classify propagation paths as \ac{LoS}, detected \ac{NLoS} from tracked targets, and non-detected \ac{NLoS} from unknown scatterers, and further explain these \ac{NLoS} components and how leveraging target information reduces complexity and overhead later in the paper.
The relative contribution of these components depends on factors such as the propagation environment, the number of targets, and their distances.
To explicitly characterize the relative contributions of these components, we decompose the channel as:
\begin{align}\label{comm_ch}
    &  {\mathbf{h}}_{k,n}(t)= \gamma \mathbf{h}^{\text{LoS}}_{k,n}(t) + \sqrt{1-\gamma^2}\Big[\tilde{\gamma} \mathbf{h}^{\text{NLoS}}_{k,n}(t) + \sqrt{1-\tilde{\gamma}^2}\mathbf{h}^R_{k,n}\Big],
\end{align}
 Equation~\eqref{comm_ch} expresses a hierarchical decomposition of the channel for user $k$ in time slot $n$. The channel is modeled as a combination of a \ac{LoS} component $\mathbf{h}^{\text{LoS}}_{k,n}(t)$ and \ac{NLoS} components. The parameter $\gamma \in [0,1]$ determines the relative strength of the \ac{LoS} component versus all \ac{NLoS} components, analogous to the Rician K-factor. To further refine the model, we introduce a second parameter $\tilde{\gamma} \in [0,1]$, which analogously specifies the relative power of the detected NLoS paths $\mathbf{h}^{\text{NLoS}}_{k,n}(t)$ compared to residual, undetected \ac{NLoS} paths $\mathbf{h}^R_{k,n}$. This hierarchical formulation allows explicit modeling of how prior knowledge of the environment, such as detected scatterers, improves channel estimation.

The tracking module provides the positions of vehicles at time slot $n$ and predicts their positions for the next time slot $n+1$ using a kinematic motion model and the estimated \ac{AoA} and range. 
The proposed channel modeling module leverages this location information to characterize the mutual influence of vehicles on received signals, treating them as potential mobile reflectors.
Following \eqref{comm_ch}, the \ac{LoS} and the detected potential \ac{NLoS} channels are related to channel parameters as:
\begin{align}\label{channel_LoS}
\mathbf{h}^{\text{LoS}}_{k,n}(t) \,\propto 
\sqrt{\alpha_{k,n}} e^{j\Phi_{k,n}} e^{j2\pi\nu_{k,n}t}  
\mathbf{a}_{(\theta_{k,n})},
\end{align}
\begin{align}\label{channel_NLoS}
\mathbf{h}^{\text{NLoS}}_{k,n}(t) &\,\propto  \sum_{i\in \mathcal{K}, i\neq k} \nu_{\text{RCS}}
\sqrt{\alpha_{i,n}\beta_{i,k,n}} e^{j\Phi_{i,k,n}}\\
 &\qquad\quad \times e^{j2\pi(\nu_{i,n}+\mu_{i,k,n})t}e^{j\pi\frac{d_0}{\lambda}\cos{\varphi_{i,k,n}}} 
\mathbf{a}_{(\theta_{i,n})}, \nonumber
\end{align}
Here, $\alpha_{k,n}={\alpha_0}(\frac{d_{k,n}}{d_0})^{-\zeta}$ and $\beta_{i,k,n}={\alpha_0}(\frac{\tilde{d}_{i,k,n}}{d_0})^{-\zeta}$ define path
loss coefficients between \ac{DFRC}-\ac{BS} with vehicle $k$ and vehicle $i$ with vehicle $k$, respectively. $ \alpha_0 $ represents the path loss at a reference distance $d_0$, $\nu_{\text{RCS}}$ indicates the reflection gain, and $\zeta$ is
the path loss exponent characterizing the signal decay with
distance. 
We highlight that $d_{k,n}$ and $\nu_{k,n} $ determine the distance and Doppler frequency between \ac{DFRC}-\ac{BS} and vehicle $k$. In contrast, $\tilde{d}_{i,k,n}$ and $ \mu_{i,k,n}$ represent the distance and Doppler frequency between vehicle $i$ and vehicle $k$, respectively. Also, $\varphi_{i,k,n}$ defines the angle \ac{AoA} at the vehicle $k$ in the signal reflected from vehicle $i$.
The absolute phases $\Phi_{k,n}$ and $\Phi_{i,k,n}$ are assumed to be known at the transmitter. However, in practice, due to feedback impairments, user mobility, and hardware limitations, the absolute phase is typically unknown at the transmitter \cite{mohammadian2021rf,nguyen1989phaseambiguity}. Therefore, in this paper, we analyze both coherent and non-coherent optimization, accounting for the random nature of the absolute phase.

The estimated channel at each time slot $n$ is derived from the estimated parameter set $ \Delta_n=\{{\hat{\theta}_{k,n},  \hat{\nu}_{k,n}},\hat{d}_{k,n}\} $, obtained through radar sensing during that time slot. 
To enable channel prediction, which leads to more accurate sensing and data transmission in the next slot, we apply the same mapping function to transform the predicted geometry into a predicted channel representation.
 Specifically, the predicted channel at the time slot $n+1$ is expressed as: 
 \begin{align}\label{comm_ch_pred}
\check{\mathbf{h}}_{k,n+1|n}( \Delta_{n+1|n})=\gamma \check{\mathbf{h}}^{\text{LoS}}_{k,n} + \sqrt{1-\gamma^2} \tilde{\gamma} \check{\mathbf{h}}^{\text{NLoS}}_{k,n},
 \end{align}
where $ \Delta_{n+1|n}=\{{\check{\theta}_{k,n+1|n},  \check{\nu}_{k,n+1|n}},\check{d}_{k,n+1|n}\} $ where the notation $\{n+1|n\}$  indicates that these values are predicted for time slot $n+1$ based on the observations and estimations from time slot $n$.   
By leveraging ISAC-based predictions of user positions, angles, and velocities, together with Doppler compensation for dominant paths\footnote{The user receiver can exploit knowledge of its own velocity to perform Doppler compensation for the dominant propagation components \cite{liu2020radarassisted}.}, the effective channel can be treated as quasi-static within each slot \cite{liu2020radarassisted}, while residual variations from unknown or non-detectable scatterers are captured in the stochastic component of our proposed model.
Accordingly, at each time slot $n$, the optimization module utilizes the predicted channel $\check{\mathbf{h}}_{k,n+1|n}( \Delta_{n+1|n})$ from the previous time slot $n-1$. 
To simplify the notation and improve readability throughout the rest of the paper, we omit the prediction subscripts and instead use $ {\mathbf{h}}_{k,n}$ and $ \Delta_n=\{{{\theta}_{k,n},  {\nu}_{k,n}},{d}_{k,n}\}$ to denote these predicted values.

\subsection{Optimization Module}
Following subsection \ref{rate_com} for downlink data transmission, we incorporate the channel model described in \eqref{comm_ch} based on the predicted values obtained from the channel modeling module into the achievable data rate expression.
We aim to design an optimization problem based on the predicted channels that ensures high sensing quality and guarantees \ac{QoS} for communication in the next time slot. We consider the radar received signal explained in subsection \eqref{radar_m} to analyze the \ac{CRB} as our sensing quality metric. 
Wang et al. \cite{wang2024cramer} derive the \ac{CRB} for a pointed target scenario in Appendix C. Following the same approach, we obtain the \ac{CRB} for estimating \( \theta_{k,n} \) as:
\begin{figure*}[!t]
\hrulefill
\begin{equation}\label{crb}
\hspace{-3mm}
	\zeta_{(\theta_{k,n})}= \frac{\sigma^2_{k,n}}{2|\Omega_{k,n}|^2N_r\delta_t}\bigg[\kappa_{k,n} \mathbf{a}^H(\mathbf{\theta_{k,n}}) \mathbf{R}_x\mathbf{a}(\mathbf{\theta_{k,n}}) + \dot{ \mathbf{a}}^{H}(\mathbf{\theta_{k,n}}) \mathbf{R}_x\dot{ \mathbf{a}}(\mathbf{\theta_{k,n}}) - 
    \frac{ \dot{ \mathbf{a}}^{H}(\mathbf{\theta_{k,n}}) \mathbf{R}_x\mathbf{a}(\mathbf{\theta_{k,n}}) + \mathbf{a}^{H}(\mathbf{\theta_{k,n}}) \mathbf{R}_x\dot{ \mathbf{a}}(\mathbf{\theta_{k,n}})}{4\mathbf{a}^H(\mathbf{\theta_{k,n}}) \mathbf{R}_x\mathbf{a}(\mathbf{\theta_{k,n}})}\bigg]^{-1}\hspace{-3mm}.\hspace{-3mm}
\end{equation}
\hrulefill
\end{figure*}
where   $\kappa_{k,n}=\pi^2(N_r^2-1)\sin^2\theta_{k,n}/12$ and $  \dot{\mathbf{a}}(\mathbf{\theta_{k,n}}) $ is the derivation of $  {\mathbf{a}}(\mathbf{\theta_{k,n}}) $. Moreover, $\mathbf{R}_x=\sum_{k=1}^{K}\hat{\mathbf{s}}_{k,n}(t)\hat{\mathbf{s}}^H_{k,n}(t)$ where $ \hat{\mathbf{s}}_{k,n}(t)= \mathbf{b}_{k,n}\circ\mathbf{w}_{k,n}s_{k,n}(t)$.
 We especially focus on robust \ac{CRB} optimization that avoids poor estimation performance. 
 Therefore, we propose the following optimization problem:

\begin{align}
\mathcal{P}_{1}:& \mathop {\rm{min}} \limits_{\mathbf{w}_{k,n}} \mathop {\rm{max}} \limits_{\theta_{i,n}, i\in\mathcal{K}}
   \: \zeta_{\theta_{i,n}}\nonumber\\
	\text{s.t.} ~~
&\text{C}1: R_{k,n} \geq C^{\text{th}}_{k,n}, ~~\forall k, n,\nonumber\\
%&\text{C}2:  \text{Tr} \big (  \mathbf{H}_{k,n}  \mathbf{w}_{k,n} \mathbf{w}^H_{k,n}  \mathbf{H}^H_{k,n}   \big ) \geq P^{\text{th}}_{\text{min}},\nonumber\\
&\text{C}2:\sum_{k=1}^{K} \|\mathbf{w}_{k,n}\|^2\leq p^{\text{lim}}_{\text{max}},~~ \forall n.\nonumber
% &\text{C}3:\sum_{k'\in \mathcal{K}, k'\not= k}\text{Tr} \big (  \mathbf{H}_{k',n}  \mathbf{w}_{k,n} \mathbf{w}^H_{k,n}  \mathbf{H}^H_{k',n}   \big ) \leq P^{\text{int}}_{\text{max}}, \forall k, n, \nonumber\\
\end{align}
We realize robust beamforming with the min-max strategy to guarantee \ac{CRB} for the worst \ac{AoA} estimation. The max-min method achieves the best
fairness in the sensing procedure. Conversely, C1 ensures that the achievable data rate for user 
$k$ in time slot 
$n$ meets a minimum threshold, denoted by $C^{\text{th}}_{k,n} $, which represents the required minimum data rate for reliable communication. As we consider the same signal for transmission and sensing, we consider C2 which ensures a limitation over the power budget for both tasks.
% We emphasize that in $\mathcal{P}_{1}$, we have $\mathbf{h}_{k,n}=\check{\mathbf{h}}_{k,n|n-1}$, which is obtained from the channel modeling module in the previous time slot. Additionally, 
%$\mathbf{H}^H_{k,n}$ is a function of the predicted angle and distance of the vehicle from the previous time slot.

\subsubsection{Solution Methodology}
Problem $\mathcal{P}_{1}$ is a min-max optimization problem that is computationally complex to solve due to the non-convex nature of its objective function and constraint C1.
To solve the min-max problem, we convert the min-max into a more tractable form by applying the epigraph method. First, with introducing the auxiliary variable $\rho_{n}$ the optimization problem $\mathcal{P}_{1}$ is reformulated as:
\begin{align}
\mathcal{P}_{2}:& \mathop {\rm{min}} \limits_{\mathbf{w}_{k,n}, \rho_{n}} 
   \: \rho_{n} \nonumber\\
	\text{s.t.} ~~
&\text{C}1: R_{k,n} \geq C^{\text{th}}_{k,n}, ~~\forall k, n,\nonumber\\
&\text{C}2:\sum_{k=1}^{K} \|\mathbf{w}_{k,n}\|^2\leq p^{\text{lim}}_{\text{max}}, ~~\forall n,\nonumber\\
& \text{C}3:  \zeta_{\theta_{k,n}}\leq \rho_{n}, ~~\forall k, n. \nonumber
\end{align}
Due to the non-convex constraints C1 and C3, the optimization problem $\mathcal{P}_{2}$ is also non-convex. To solve C1, we consider $\Gamma_{k,n}\geq e^{C^{\text{th}}_{k,n}}-1$. Hence, we have 
\begin{align}\label{C_SINR_t}
&\Dot{C}1: \left |\sqrt{N_t}\mathbf{h}_{k,n}^H\mathbf{w}_{k,n}\right |^2 \nonumber\\
& \geq (\underset{i\in\mathcal{K}\setminus \left \{ k \right \}}{\sum }\left |\sqrt{N_t}\mathbf{h}_{k,n}^H\mathbf{w}_{i,n}\right |^2 +  \sigma^2_{k,n})(e^{C^{\text{th}}_{k,n}}-1),~~\forall k, n.
\end{align} 
 Since $ e^{C^{\text{th}}_{k,n}}-1\geq 0 $ for the positive values of $ C^{\text{th}} $, both the left-hand side and right-hand side of constraint $\Dot{C}1$ are convex functions of the beamforming vector. Therefore, $\Dot{C}1$ is a non-convex constraint. To address this non-convexity, we apply the following Taylor approximation to facilitate its solution:
\begin{align}\label{C_SINR_t_1}
 &\left |\sqrt{N_t}\mathbf{h}_{k,n}^H\mathbf{w}_{k,n}\right |^2\approx 2N_t \mathcal{R}\bigg( (\mathbf{h}_{k,n}^H {\mathbf{w}_{k,n}}^{[j]})^{H}\mathbf{h}_{k,n}^H \mathbf{w}_{k,n}\bigg)\nonumber \\
 & - N_t{\left|\mathbf{h}_{k,n}^H {\mathbf{w}_{k,n}}^{[j]} \right|}^2,~~\forall k, n. 
\end{align}
%\begin{align}\label{C_SINR_t}
%\left|\sqrt{N_t}\mathbf{h}_{k,n}^H\mathbf{w}_{k,n}\right|^2 
%&\approx 2 N_t \, \mathcal{R} \Big( (\mathbf{h}_{k,n}^H {\mathbf{w}_{k,n}}^{[j]})^H 
%\mathbf{h}_{k,n}^H \mathbf{w}_{k,n} \Big) \nonumber\\
%&\phantom{\approx} - N_t \, \left|\mathbf{h}_{k,n}^H {\mathbf{w}_{k,n}}^{[j]} \right|^2
%\end{align}
Hence, constraint $ \Dot{C}1 $ can be approximated by the following convex constraint:
\begin{align}\label{C_SINR_t_2}
&\tilde{C}1: 2N_t \mathcal{R}\bigg( (\mathbf{h}_{k,n}^H {\mathbf{w}_{k,n}}^{[j]})^{H}\mathbf{h}_{k,n}^H \mathbf{w}_{k,n}\bigg)-N_t{\left|\mathbf{h}_{k,n}^H {\mathbf{w}_{k,n}}^{[j]} \right|}^2 \nonumber \\
& \geq (\underset{i\in\mathcal{K}\setminus \left \{ k \right \}}{\sum }\left |\sqrt{N_t}\mathbf{h}_{k,n}^H\mathbf{w}_{i,n}\right |^2+  \sigma^2_{k,n})(e^{C^{\text{th}}_{k,n}}-1),~~\forall k, n.
\end{align}
On the other side, introducing the new slack variable $t_{k,n}$, we use the Schur complement  condition to reformulate C3 with the following constraints:

\begin{subequations} \label{crb_f_ff}
\begin{align}
& \Dot{\text{C}}3: \:   \dfrac{ \sigma_{k,n}^2 }{2|\Omega_{k,n}|^2N_r\delta_t}\leq \rho_{n} t_{k,n},~~ \forall k,n,\\
& \Ddot{\text{C}}3: \: {\begin{bmatrix}
\kappa_{k,n}{A}_{k,n}+{B}_{k,n}-t_{k,n} & {E}_{k,n}\\
{F}_{k,n} & 4{A}_{k,n}
\end{bmatrix}}\succcurlyeq \textbf{0}, ~~\forall k, n. 
\end{align}
\end{subequations} 
Functions ${A}_{k,n}$, ${B}_{k,n}$, ${E}_{k,n}$, and ${F}_{k,n}$ are defined as follows:
\begin{subequations} \label{crb_f_details}
\begin{align}
& {A}_{k,n}=\mathbf{a}^H(\mathbf{\theta_{k,n}}) \mathbf{R}_x\mathbf{a}(\mathbf{\theta_{k,n}}), ~~\forall k,n,\nonumber\\
& {B}_{k,n}=\dot{ \mathbf{a}}^{H}(\mathbf{\theta_{k,n}}) \mathbf{R}_x\dot{ \mathbf{a}}(\mathbf{\theta_{k,n}}), ~~\forall k,n,\nonumber\\
& {E}_{k,n}=\dot{ \mathbf{a}}^{H}(\mathbf{\theta_{k,n}}) \mathbf{R}_x\mathbf{a}(\mathbf{\theta_{k,n}}),~~ \forall k,n,\nonumber\\
& {F}_{k,n}=\mathbf{a}^{H}(\mathbf{\theta_{k,n}}) \mathbf{R}_x\dot{ \mathbf{a}}(\mathbf{\theta_{k,n}}), ~~\forall k,n,\nonumber
\end{align}
\end{subequations}
Constraint $ \Dot{\text{C}}3$ is still non-convex due to the multiplication of two variables $\rho_{n}$ and $t_{k,n}$. Hence, we utilize the \ac{SCA} approach as follows:
\begin{align}\label{r_t_app}
&{\rho_{n} t_{k,n}}\approx \rho_{n}^{[j]}t_{k,n}+\rho_{n}^{[j]}(t_{k,n}-t_{k,n}^{[j]})\nonumber\\
& +t_{k,n}^{[j]}(\rho_{n} -\rho_{n}^{[j]})\triangleq\chi_{k,n},~~\forall k, n  .
\end{align}
On the other side, to convexify $\Ddot{\text{C}}3 $, we find the surrogate functions of ${A}_{k,n}$, ${B}_{k,n}$, ${E}_{k,n}$, and ${F}_{k,n}$ around $ \tilde{\mathbf{w}}_{i,n}^{[j]}= \mathbf{b}_{i,n}\circ{\mathbf{w}_{i,n}}^{[j]}$ as follows: 
\begin{subequations} \label{crb_f_details_1}
\begin{align}
& \tilde{{A}}_{k,n}=\sum_{i=1}^{K}2\mathcal{R}\bigg( (\mathbf{a}^H(\mathbf{\theta_{k,n}}) {\tilde{\mathbf{w}}_{i,n}}^{[j]})^{H}\mathbf{a}^H(\mathbf{\theta_{k,n}}) \tilde{\mathbf{w}}_{i,n}\bigg)\nonumber\\
& -{\left|\mathbf{a}^H(\mathbf{\theta_{k,n}}) {\tilde{\mathbf{w}}_{i,n}}^{[j]} \right|}^2, ~~\forall k,n,\nonumber\\
& \tilde{{B}}_{k,n}=\sum_{i=1}^{K}2\mathcal{R}\bigg( (\Dot{\mathbf{a}}^H(\mathbf{\theta_{k,n}}) {\tilde{\mathbf{w}}_{i,n}}^{[j]})^{H}\Dot{\mathbf{a}}^H(\mathbf{\theta_{k,n}}) \tilde{\mathbf{w}}_{i,n}\bigg)\nonumber\\
& -{\left|\Dot{\mathbf{a}}^H(\mathbf{\theta_{k,n}}) {\tilde{\mathbf{w}}_{i,n}}^{[j]} \right|}^2,~~ \forall k,n,\nonumber\\
& \tilde{{E}}_{k,n}=\sum_{i=1}^{K}\bigg( \Dot{\mathbf{a}}(\mathbf{\theta_{k,n}}){\mathbf{a}}^H(\mathbf{\theta_{k,n}}) +{\mathbf{a}}(\mathbf{\theta_{k,n}})\Dot{\mathbf{a}}^H(\mathbf{\theta_{k,n}})\bigg)\nonumber\\
&  \tilde{\mathbf{w}}_{i,n}^{[j]}(\tilde{\mathbf{w}}_{i,n}-\tilde{\mathbf{w}}_{i,n}^{[j]})+{E}_{i,n}^{[j]},~~ \forall k,n,\nonumber\\
& \tilde{{F}}_{k,n}=\sum_{i=1}^{K}\bigg( {\mathbf{a}}(\mathbf{\theta_{k,n}})\Dot{\mathbf{a}}^H(\mathbf{\theta_{k,n}}) +\Dot{\mathbf{a}}(\mathbf{\theta_{k,n}}){\mathbf{a}}^H(\mathbf{\theta_{k,n}})\bigg) \nonumber\\
&  \tilde{\mathbf{w}}_{i,n}^{[j]}(\tilde{\mathbf{w}}_{i,n}-\tilde{\mathbf{w}}_{i,n}^{[j]})+{F}_{i,n}^{[j]}, ~~\forall k,n.\nonumber
\end{align}
\end{subequations}
Based on \eqref{r_t_app} and the introduced definitions of $\tilde{{A}}_{k,n}$, $\tilde{{B}}_{k,n}$, $\tilde{{E}}_{k,n}$, and $\tilde{{F}}_{k,n}$, 
we replace constraints $ \Dot{C}3$ and $ \Ddot{C}3$ with the following constraints: 
\begin{subequations} \label{crb_f}
\begin{align}
& \hat{\text{C}}3: \:   \dfrac{ \sigma_{k,n}^2 }{2|\Omega_{k,n}|^2N_r\delta_t}\leq \chi_{k,n}, ~~\forall k,n,\\
& \tilde{\text{C}}3: \: {\begin{bmatrix}
\kappa_{k,n}\tilde{{A}}_{k,n}+\tilde{{B}}_{k,n}-t_{k,n} & \tilde{{E}}_{k,n}\\
\tilde{{F}}_{k,n} & 4\tilde{{A}}_{k,n}
\end{bmatrix}}\succcurlyeq \textbf{0}, ~~\forall k, n. 
\end{align}
\end{subequations}
As a result, we reformulate the problem $ \mathcal{P}_{2}$ as follows:
\begin{align}
\mathcal{P}_{3}:& \mathop {\rm{min}} \limits_{\mathbf{w}_{k,n},\: \rho_{n},\: t_{k,n}} 
   \: \rho_{n} \nonumber\\
	\text{s.t.} ~~
&\tilde{\text{C}}1, \text{C}2, \hat{\text{C}}3, \tilde{\text{C}}3. \nonumber
\end{align}
Problem $\mathcal{P}_{3}$ is a convex problem that jointly optimizes variables $\mathbf{w}_{k,n}$, $\rho_{n}$, and $t_{k,n}$. To solve this problem, we consider an iterative algorithm following the \ac{SCA} approach. Hence, in each iteration $j$, we update the solution set and efficiently solve problem $\mathcal{P}_{3}$ via CVX. To provide a simplified overview of the proposed framework, including the optimization solution, Algorithm~\ref{alg1} summarizes its main steps.

We highlight that in \ac{mmWave} networks, numerous studies indicate that \ac{mmWave} performance is highly affected by penetration losses and blockages, reducing coverage and reliability~\cite{raghavan2019statistical}. Hence, we incorporate a blockage detection mechanism using received backscattered echoes. By leveraging the detected scatterers, we enhance network performance even in the presence of blockages.

\begin{algorithm}[t]
\caption{Proposed \ac{ISAC} Framework Including Resource Allocation}
\label{alg1}

\textbf{Input:} Initial positions and velocities of the vehicles, total number of time slots $N$, and iteration index $j$.

\textbf{Output:} Optimized beamforming vectors, slack variables, and epigraph-based fairness variable
$({\textbf{w}^*_{k,n}}, {t}^*_{k,n}, {\rho^*_{n}})$.

\vspace{0.5em}

\textbf{For} $n = 1, \dots, N$ \textbf{do}

\begin{enumerate}
    \item \textbf{Detection:} Apply \ac{APES} filter to detect objects.

    \item \textbf{Initialization:} Set iteration counter $j = 0$.

    \item \textbf{Iterative optimization:}
    \begin{itemize}
        \item Solve $\mathcal{P}_3$ to obtain
        ${\textbf{w}^{[j+1]}_{k,n}}, {t}^{[j+1]}_{k,n}, {\rho^{[j+1]}_{n}}$.
        \item Update iteration counter: $j \leftarrow j + 1$.
    \end{itemize}

    \item \textbf{Convergence check:}
    \[
    \frac{\mathcal{O}bj^{[j]} - \mathcal{O}bj^{[j-1]}}{\mathcal{O}bj^{[j-1]}} \leq \epsilon
    \]

    \item \textbf{Return optimization result:}
    $({\textbf{w}^*_{k,n}}, {t}^*_{k,n}, {\rho^*_{n}})$

    \item \textbf{Estimation:} Analyze received backscattered signals using \ac{MF} and \ac{MUSIC} to obtain
    $\tau_{k,n}, \nu_{k,n}, \theta_{k,n}$.

    \item \textbf{\ac{EKF}:} Refine estimated parameters using \ac{EKF} \cite{liu2020radarassisted}.

    \item \textbf{Prediction:} Predict next state:
    $\check{\theta}_{k,n}, \check{\tau}_{k,n}, \check{\nu}_{k,n}$.

    \item \textbf{Channel reconstruction:} Compute channel using
    \eqref{channel_LoS}, \eqref{channel_NLoS}, and \eqref{comm_ch}.
\end{enumerate}

\textbf{End For}

\end{algorithm}

\subsubsection{Robust non-coherent optimization}
Although the target parameters can be estimated with high accuracy, the absolute phase information remains subject to significant uncertainty. This arises from multiple factors, including synchronization mismatches between the transmitter and receiver, delayed or impaired feedback channels, hardware imperfections such as phase noise or quantization errors, and the presence of unknown or unmodeled scatterers in the propagation environment \cite{moghaddam2023statistical,guo2025integrated}.
The absolute phase of multipath components is directly related to fast fading, which arises from rapid constructive and destructive interference. This effect is especially pronounced in \ac{mmWave} scenarios, where movements of just a few millimeters can cause significant phase shifts, resulting in rapid fluctuations of the absolute phase with user motion \cite{snchez2021fading}.
These factors collectively contribute to residual phase errors that cannot be eliminated through conventional estimation techniques. 
While coherent estimation relies on precise phase knowledge, such information is not available in non-coherent estimation, necessitating the use of alternative methods.
To rigorously account for this uncertainty in the system model, we treat the absolute phase as a random variable, denoted by $\Phi_{k,n}$, and assume it follows a uniform distribution over the interval $[0,2\pi)$, i.e., $\Phi_{k,n} \sim \mathcal{U}[0,2\pi)$. This modeling approach allows the framework to capture the full range of possible phase deviations and ensures that subsequent analyses of channel estimation, beamforming, and sensing performance incorporate the impact of phase uncertainties. As a result, we define a new optimization problem to make sure of the achievable data rate for all possible absolute phases as follows:
\begin{align}
\mathcal{P}_{4}:& \mathop {\rm{min}} \limits_{\mathbf{w}_{k,n}} \mathop {\rm{max}} \limits_{\theta_{i,n}, i\in\mathcal{K}}
   \: \zeta_{\theta_{i,n}}\nonumber\\
	\text{s.t.} ~~
&\Hat{\text{C}}1: R_{k,n}(\Phi_{k,n}) \geq C^{\text{th}}_{k,n}, \: \Phi_{k,n} \sim \mathcal{U} \: [0,2\pi),\: \forall k, n, \nonumber\\
&\text{C}2:\sum_{k=1}^{K} \|\mathbf{w}_{k,n}\|^2\leq p^{\text{lim}}_{\text{max}}, ~~\forall n.\nonumber
% &\text{C}3:\sum_{k'\in \mathcal{K}, k'\not= k}\text{Tr} \big (  \mathbf{H}_{k',n}  \mathbf{w}_{k,n} \mathbf{w}^H_{k,n}  \mathbf{H}^H_{k',n}   \big ) \leq P^{\text{int}}_{\text{max}}, \forall k, n, \nonumber\\
\end{align}
In problem $\mathcal{P}_{4}$, constraint $\Hat{\text{C}}1$ ensures that achievable data rates are satisfied for all possible values of the random phase parameter $\Phi_{k,n}$. To solve this problem, we follow the solution methodology introduced for $\mathcal{P}_{1}$, with a modification to account for the stochastic nature of $\Hat{\text{C}}1$. Specifically, we employ a \ac{SAA} approach, drawing $U$ independent samples ${\Phi^1_{k,n}, \dots, \Phi^U_{k,n}}$ from the distribution of $\Phi_{k,n}$. Using our \ac{SCA}-based approximation, the convex reformulation of $\Hat{\text{C}}1$ is given by:
\begin{align}\label{C_SINR_t_3}
&\bar{C}1: 2N_t \mathcal{R}\bigg( (\mathbf{h}_{k,n}^H(\Phi_{k,n}^u) {\mathbf{w}_{k,n}}^{[j]})^{H}\mathbf{h}_{k,n}^H(\Phi_{k,n}^u) \mathbf{w}_{k,n}\bigg)\nonumber \\
&-N_t{\left|\mathbf{h}_{k,n}^H(\Phi_{k,n}^u) {\mathbf{w}_{k,n}}^{[j]} \right|}^2 \nonumber \\
& \geq (\underset{i\in\mathcal{K}\setminus \left \{ k \right \}}{\sum }\left |\sqrt{N_t}\mathbf{h}_{k,n}^H(\Phi_{k,n}^u)\mathbf{w}_{i,n}\right |^2+  \sigma^2_{k,n})(e^{C^{\text{th}}_{k,n}}-1).
\end{align}
Using these samples, the resulting deterministic convex problem can be formulated as follows:
\begin{align}
\mathcal{P}_{5}:& \mathop {\rm{min}} \limits_{\mathbf{w}_{k,n},\: \rho_{n},\: t_{k,n}} 
   \: \rho_{n} \nonumber\\
	\text{s.t.} ~~
&\bar{\text{C}}1, \text{C}2, \hat{\text{C}}3, \tilde{\text{C}}3. \nonumber
\end{align}
 It is worth noting that increasing the number of samples $U$ improves the accuracy of the approximation and better captures the impact of phase uncertainty on the constraint.

\subsection{ Overall Algorithm}
Algorithm~1 presents the AO-based solution, and the following sections analyze its convergence properties and computational complexity.
\subsubsection{Convergence Analysis}
In the proposed \ac{SCA}-based algorithm, the original nonconvex problem is approximated
by a convex surrogate at each iteration $j$ using first-order Taylor expansions. 
The convergence follows from standard \ac{SCA} arguments based on feasibility, 
monotonicity, and boundedness \cite{scutari2018parallel}. Specifically, each iterate 
$(\mathbf{w}_{k,n}^{[j]}, \rho_n^{[j]}, t_{k,n}^{[j]})$ 
is feasible for the original problem, since the convex approximations form inner approximations
of the original constraints. Moreover, as $\mathcal{O}bj$ is minimized at each iteration, 
the sequence of objective values is non-increasing:
\[
\mathcal{O}bj^{[j+1]} \le \mathcal{O}bj^{[j]}.
\]
Since $\rho_n$ is lower bounded (e.g., $\rho_n \ge 0$), the sequence 
$\{\rho_n^{[j]}\}$ converges to a finite value. Therefore, the algorithm converges 
to a stationary point $(\mathbf{w}_{k,n}^*, \rho_n^*, t_{k,n}^*)$ of the original problem. 
Although the objective depends only on $\rho_n$, the auxiliary variable $t_{k,n}$ 
and the main variable $\mathbf{w}_{k,n}$ are optimized simultaneously to satisfy 
the constraints, preserving the convergence guarantees of \ac{SCA}.
\subsubsection{Computational Complexity}
Here, we analyze the computational complexity of Algorithm~\ref{alg1} based on the \ac{SCA} method. The computational complexity is given by
$\mathcal{O}\Big(\mathrm{log}(1/\epsilon)P_{\text{S}}C_{\text{S}}^3    \Big)$ where $\mathcal{O}(\cdot)$ denotes the big-O notation, and 
 $\epsilon$ represents the desired solution accuracy \cite{boyd2004convex}. Here, $P_{\text{S}}$  denotes the problem size, and $C_{\text{S}} $  is the number of constraints.
For problem $\mathcal{P}_{3}$, the problem size and number of constraints are given by $P_{\text{S}}= KN_t$ and $C_{\text{S}}= 3K+1$.
In contrast, conventional \ac{SDR}-based approaches  lift each beamforming vector into a $N_t \times N_t$ matrix, resulting in
$P_S^{\text{SDR}} = K N_t^2$ variables and $C_S^{\text{SDR}} = 4K+1$ constraints\cite{luo2010semidefinite}. 
The corresponding per-iteration complexity is $\mathcal{O}\big(P_S^{\text{SDR}} ((1/\epsilon)C_S^{\text{SDR}})^3\big) = \mathcal{O}\big((1/\epsilon)K N_t^2 (4K+1)^3\big)$.
It is evident that the proposed SCA approach significantly reduces computational cost, particularly in systems with large antenna arrays.
For problem $\mathcal{P}_{5}$,  they are $P_{\text{S}}= KN_t+K+1$ and $C_{\text{S}}= k(2+U)+1$.  
The stated complexity corresponds to a single time slot; for the entire frame, the total complexity is obtained by multiplying by the number of time slots.

%%%%%%%%%%%%%%%%%%%%%%%%%%%%%%%%%%%%%%%%%%%%%%%%%%%%%%%%%%%%%
%%   Simulation Results
%%%%%%%%%%%%%%%%%%%%%%%%%%%%%%%%%%%%%%%%%%%%%%%%%%%%%%%%%%%

\section{Simulation Results}\label{simulations}

In this section, we evaluate the efficiency of the proposed
optimization framework for ISAC, emphasizing its performance in
terms of detection and tracking accuracy, as well as communication
effectiveness. We assess the proposed framework across several
aspects, including estimation accuracy and tracking validation,
communication performance in terms of achievable data rate, blockage
detection capability, and robust beamforming under absolute phase
ambiguity.

\subsection{Simulation Scenario}
The road segment is modeled as a three-lane roadway aligned with the
$x$-axis, with lane centerlines at $y \in \{22, 25, 28\}$ m and a
lane width of $3$ m. The \ac{DFRC}-\ac{BS} is located at the origin
$(0,0)$, offset from the roadway to emulate a roadside deployment.
In the baseline scenario ($K=3$), three vehicles are initialized at
positions $O_1=(25,25)$, $O_2=(35,25)$, $O_3=(40,25)$ m, moving in the
negative $x$-direction at constant velocities $v_1=55$, $v_2=50$, and
$v_3=35$ km/h, respectively, representative of moderate urban traffic
speeds. To examine denser traffic, six additional vehicles are placed
at $O_4$--$O_9$ (Table~\ref{tab:vehicles}) with velocities ranging
from $60$--$85$ km/h, yielding $K=9$ users in total.

 To evaluate scalability beyond this baseline (\Cref{comp_rmse_ues}),
we further consider up to $K=35$ users with initial positions drawn
uniformly at random in the x-position within the road segments to represent a denser traffic configuration.
Moreover, to capture realistic non-constant vehicle motion, we additionally
consider a scenario in which vehicle velocities increase gradually
over the tracking horizon by up to $30$ km/h from their initial
values (\Cref{RMSE_CDF}), reflecting acceleration behavior and
heterogeneous speed evolution across users, rather than assuming
fixed velocities throughout.

\begin{table}[t]
\centering
\caption{Initial vehicle positions and velocities.}
\label{tab:vehicles}
\begin{tabular}{ccccc}
\toprule
Vehicle & Lane ($y$, m) & Initial position (x, m) & Velocity (km/h) \\
\midrule
$O_1$ & 25 & 25 & 55 \\
$O_2$ & 25 & 35 & 50 \\
$O_3$ & 25 & 40 & 35 \\
\midrule
$O_4$ & 28 & 29 & 70 \\
$O_5$ & 28 & 40 & 65 \\
$O_6$ & 28 & 44 & 60 \\
$O_7$ & 22 & 55 & 85 \\
$O_8$ & 22 & 50 & 80 \\
$O_9$ & 22 & 45 & 70 \\
\bottomrule
\end{tabular}
\end{table}

We consider that, in addition to the detectable scatterers (other vehicles in our case study), there are additional scatterers, such as buildings and trees, whose effects are modeled using the Rayleigh fading model with the power-splitting
coefficients $\tilde{\gamma}=0.8$.  
The detailed simulation parameters are summarized in Table \ref{ISAC_parameters}  \cite{du2023integrated} and \cite{khalili2024efficient}.
We assess the proposed framework across several aspects, including estimation accuracy and tracking validation, communication performance in terms of achievable data rate, blockage detection capability, and robust beamforming under absolute phase ambiguity.

\begin{table}[t]
    \centering
    \caption{System simulation parameters.}
    \label{ISAC_parameters}
    \begin{tabular}{lll}
    \toprule
    \textbf{Symbol} & \textbf{Parameter} & \textbf{Value} \\
    \midrule
    $\sigma^{2}, \sigma_{k,n}^2$ & Noise power & $-80$ dBm \\
    $N$ & Total number of time slots & $104$ \\
    $\delta_{t}$ & Duration of one time slot & $0.02$ s \\
    $p^{\text{lim}}_{\mathrm{max}}$ & Maximum power budget & $35$ dBm \\
    ${\Gamma}^\text{th}_{k,n}$ & Required SNR of each user & $-5$ dB \\
    $N_t$ & Number of transmit antennas & $64$ \\
    $N_r$ & Number of receive antennas & $64$ \\
    ${\zeta}$ & Path loss exponent & $2$ \\
    $f_c$ & Carrier frequency & $60$ GHz \\
    $BW$ & Bandwidth & $200$ MHz \\
    $\tilde{\gamma}$ & Power-splitting coefficient & $0.8$ \\
    \bottomrule
    \end{tabular}
\end{table}

\begin{figure} 
	\centering
	%\vspace{-3mm}
 	\subfigure[Estimation of angle with EKF.]{
		\label{ekf_theta}
		%\vspace{-3mm}
		\includegraphics[height=4.6cm]{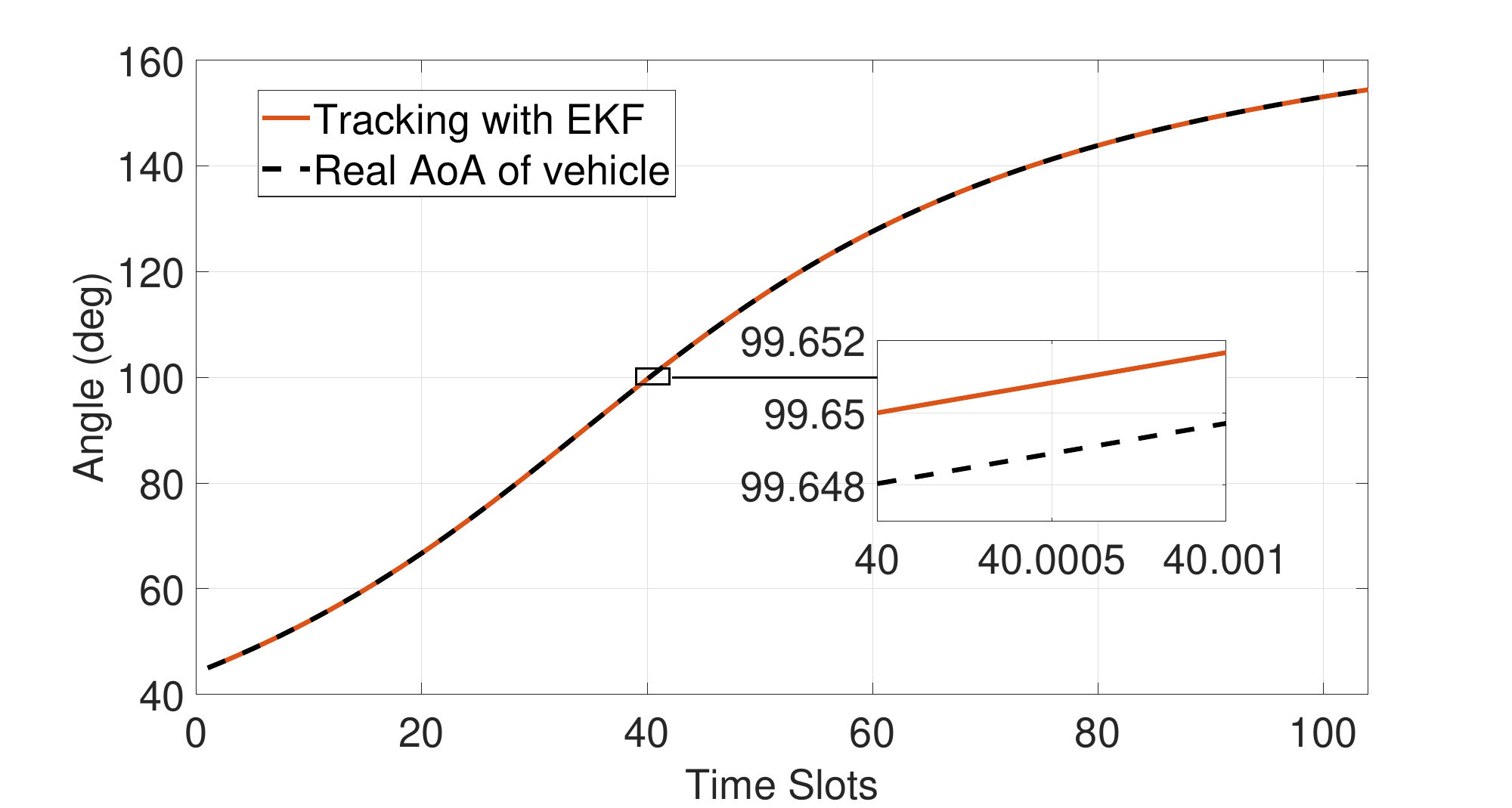}%\includegraphics[height= 6.500cm,width=7.8cm]{with_EKF_aoa_20_10_2025.pdf}
	}
	\subfigure[Estimation of angle without EKF in each time slot.]{
		\label{nekf_theta}
		\includegraphics[height=4.6cm]{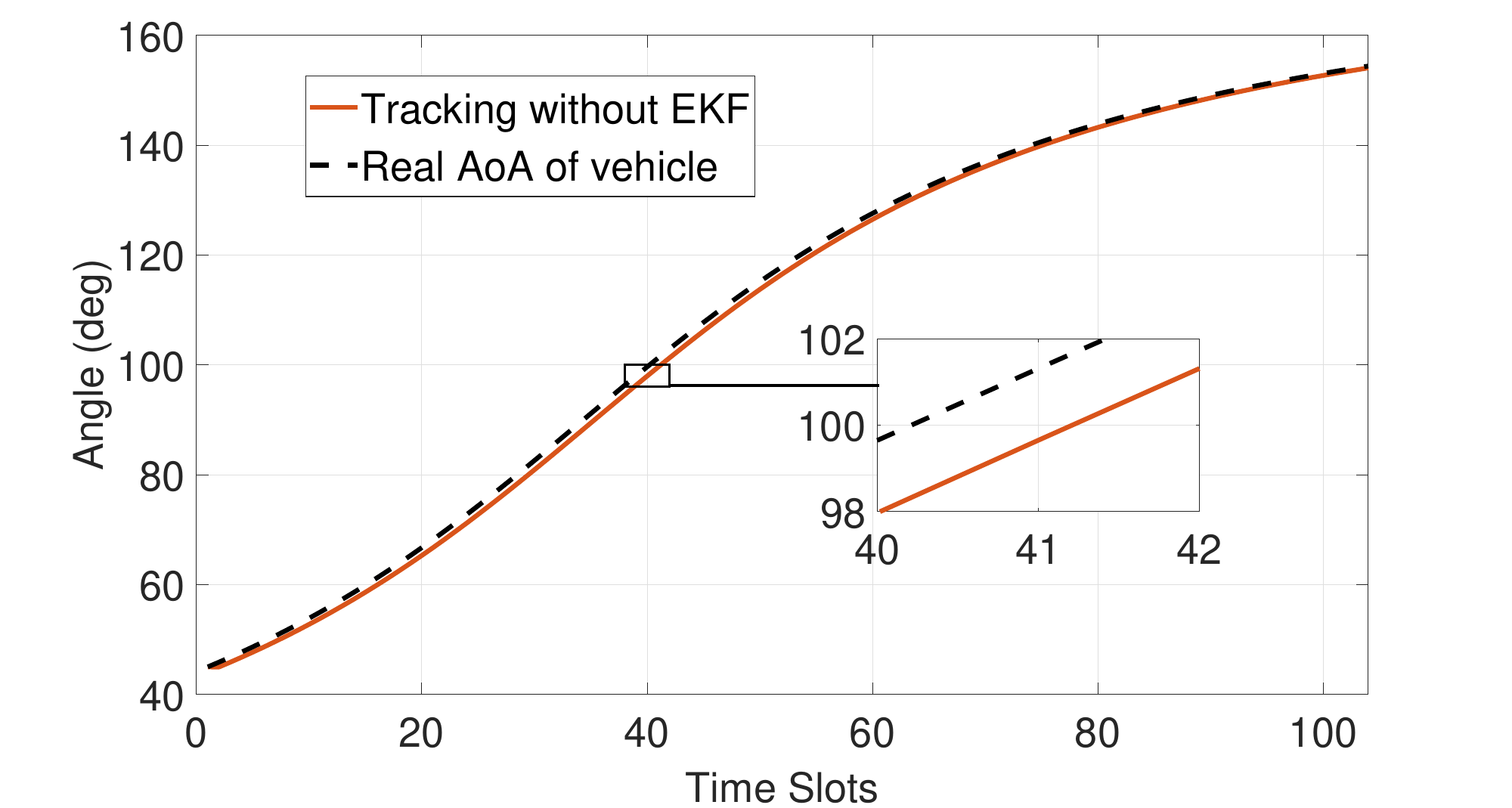}%\includegraphics[height= 6.500cm,width=7.8cm]{No_EKF_aoa_20_10_2025.pdf}
	}
    \caption {Performance analysis of state estimation accuracy for UE 1 using EKF versus a non-EKF approach.} 
\end{figure}

%----------------------------------

\begin{figure} \label{AoA_5G_NR}
	\centering
	%\vspace{-3mm}
 	\subfigure[AoA estimation accuracy of UE 1.]{
		\label{AoA_5G_NR_ue1}
		%\vspace{-3mm}
	\includegraphics[height=4.6cm]{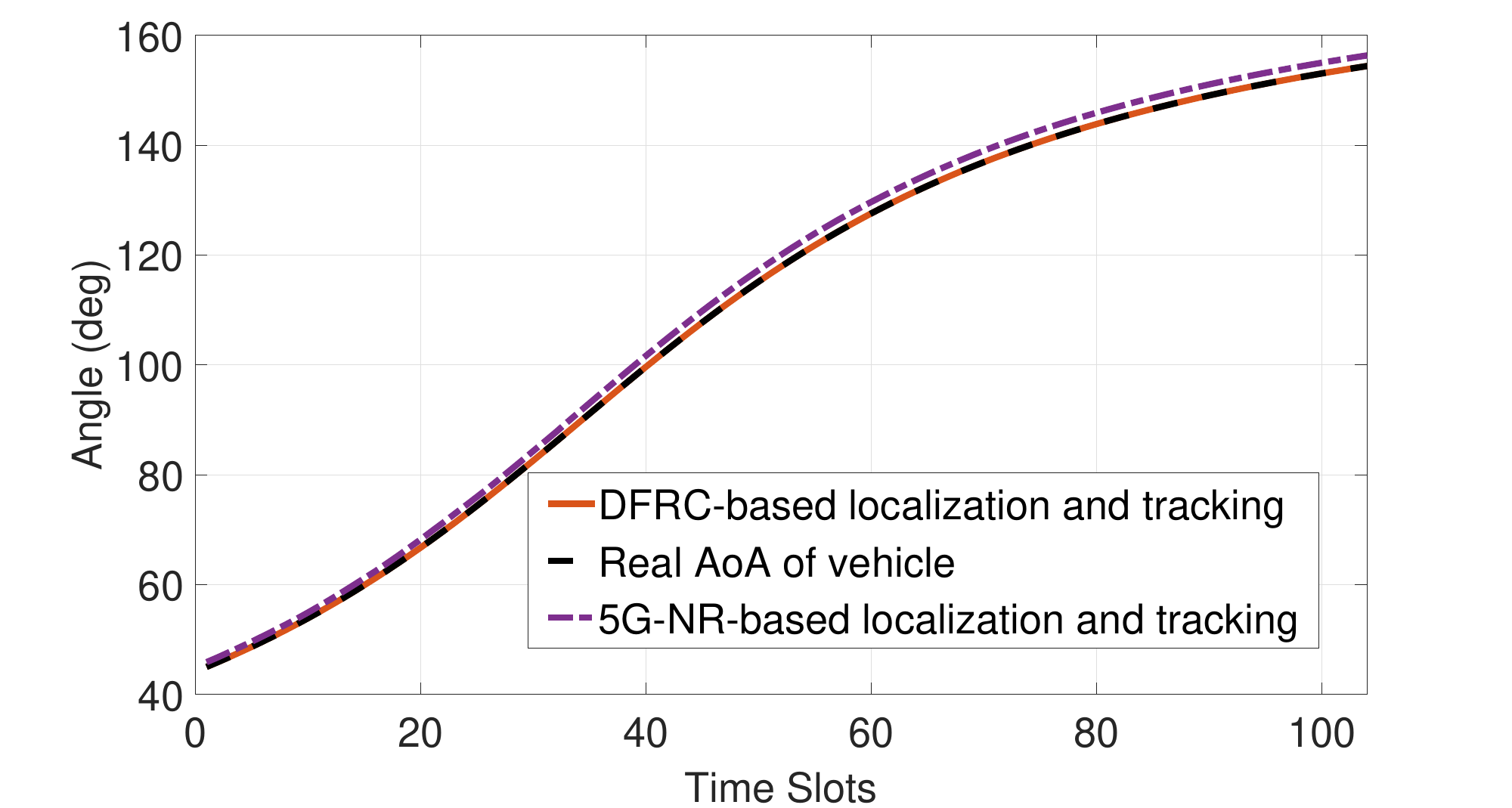}%\includegraphics[height= 6.500cm,width=7.8cm]{figures/with_EKF_aoa_20_10_2025.pdf}
	}
	\subfigure[Range estimation accuracy of UE 1.]{
		\label{distance_5G_NR_ue1}
		\includegraphics[height=4.6cm]{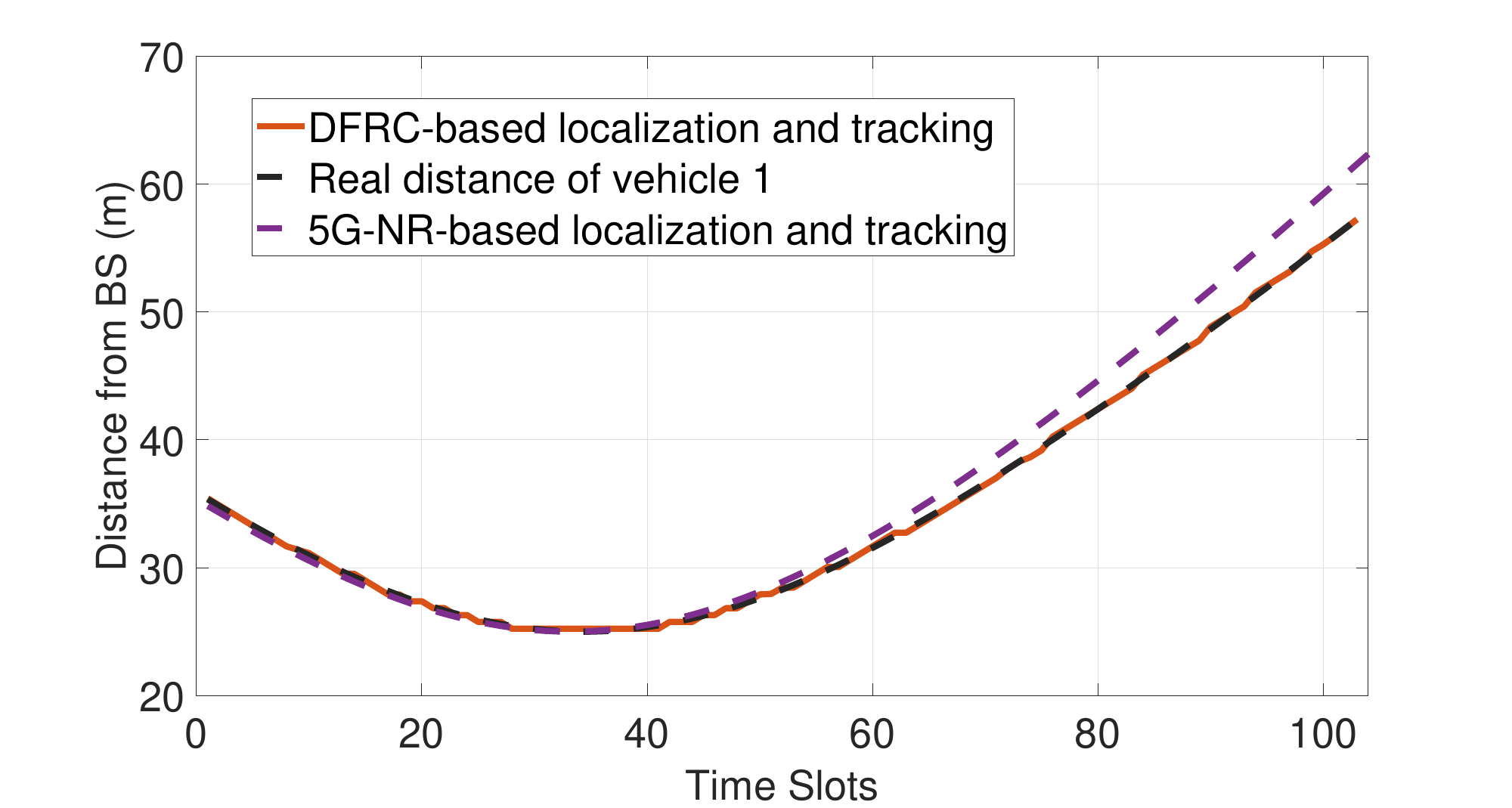}%
	}
    \caption {Accuracy Analysis Considering DFRC- and Communication-Based Tracking and Localization for UE 1.} 
\end{figure}

\subsection{Estimation accuracy and tracking validation}
\Cref{ekf_theta} and \Cref{nekf_theta} present the target tracking over the entire time horizon, demonstrating that the angle estimation remains highly accurate due to the use of high-precision methods such as the \ac{MUSIC} algorithm. In addition, \Cref{ekf_theta} illustrates the \ac{AoA} tracking of one of the vehicles using the \ac{EKF}, while \Cref{nekf_theta} presents the tracking performance without employing the \ac{EKF} gain. To be more precise, we show, for each time slot, the amount of inaccuracy that occurs if the \ac{EKF} refinement provided by the \ac{KF} gain is not applied within that slot. The comparison highlights the absence of error mitigation normally achieved through the \ac{KF} gain at each time slot. As an example, we zoom in on one of the predicted angles to emphasize this effect. 

\Cref{AoA_5G_NR_ue1} and \Cref{distance_5G_NR_ue1} clearly demonstrate the superior performance of the proposed \ac{DFRC}-based scheme compared to the communication-based localization and tracking approach for the first user. The communication-based scheme exhibits significantly lower accuracy, with performance degrading progressively as the tracking duration increases. In particular, the estimation error between the communication-based approach and the true \ac{AoA} grows over time, highlighting its limited tracking robustness. Moreover, in terms of range detection and tracking for Vehicle~1, the communication-based scheme shows substantially reduced accuracy, whereas the proposed DFRC-based method consistently maintains reliable and stable performance throughout the observation interval. Overall, considering both AoA and range estimation under tracking, the proposed DFRC-based localization framework demonstrates superior reliability and robustness.

\begin{table}[t]
\centering
\caption{Performance comparison with benchmark schemes.}
\label{tab:comparison}
\begin{tabular}{lcc}
    \toprule
    \textbf{Worst Tracking Error (RMSE)} & \textbf{Angle (°)} & \textbf{Range (m)} \\
    \midrule
    Communication-based (5G-NR)                       & 2.73   & 5.26 \\
    Separated S\&C              & 0.0284 & 1.07 \\
        Sensing-only              & 0.0094 & 0.34 \\
    Interference-unaware 
    predictive ISAC~\cite{liu2020radarassisted} & 0.0040 & 0.15 \\
    Interference-aware 
    predictive ISAC~\cite{rezaei2026mobility-aware} & 0.0183 & 0.47 \\
    \textbf{Ours}               & \textbf{0.0129} & \textbf{0.38} \\
    \bottomrule
\end{tabular}
\end{table}

To evaluate the tracking performance of the proposed ISAC framework, Table~\ref{tab:comparison} compares the worst-case angle and range RMSEs of the proposed scheme with those of five benchmark approaches: communication-based tracking, separated \ac{SC}, sensing-only, and two predictive beamforming benchmarks proposed in \cite{liu2020radarassisted} and \cite{rezaei2026mobility-aware}.The communication-based approach exhibits the largest tracking errors, with angle and range RMSEs of $2.73^\circ$ and $5.26$~m, respectively, since it does not exploit the dedicated sensing measurements available in the proposed framework. In contrast, the separated \ac{SC} scheme substantially improves tracking accuracy by allocating dedicated resources to sensing. For this baseline, we evaluate a separated sensing-and-communication architecture in which sensing and data transmission are performed independently. We further investigate the impact of the sensing duration on its tracking performance. When the sensing duration is reduced to $t_s=0.005$~s, both sensing and tracking accuracies deteriorate. This degradation is expected because a shorter sensing interval provides less received sensing information for reliable target-parameter estimation, thereby reducing the reliability of subsequent tracking updates. These results highlight the trade-off between sensing duration and tracking accuracy in a separated \ac{SC} architecture, whereas the proposed ISAC framework exploits sensing and communication within the same transmission structure while explicitly accounting for interference. 
The interference-unaware benchmark based on \cite{liu2020radarassisted} achieves the lowest tracking errors, with angle and range RMSEs of $0.0040^\circ$ and $0.15$~m, respectively. This result is expected because interference is not explicitly incorporated into its beamforming design. The proposed ISAC scheme achieves angle and range RMSEs of $0.0129^\circ$ and $0.38$~m, respectively. Although its tracking accuracy is lower than that of the interference-unaware benchmark, this degradation is expected because the proposed scheme explicitly accounts for inter-user interference while jointly satisfying the sensing and communication requirements. Thus, the proposed framework provides accurate tracking while operating under a more practical interference-aware joint sensing and communication setting. We also consider a sensing-only scheme that optimizes the beamforming design exclusively for sensing performance. This benchmark represents the tracking accuracy achievable when the available transmission resources are dedicated entirely to sensing. The sensing-only scheme achieves an angle RMSE of $0.0094^\circ$ and a range RMSE of $0.34$~m, outperforming the proposed ISAC scheme in terms of tracking accuracy. The comparison therefore highlights the trade-off between dedicated sensing performance and ISAC in the proposed framework. 
On the other hand, we consider the interference-aware scheme proposed in \cite{rezaei2026mobility-aware}, which represents one of the closest related studies to our work. As shown in Table~\ref{tab:comparison}, the proposed framework achieves higher localization accuracy compared with this interference-aware approach.

\begin{figure}[pt]
    \centering
    \includegraphics[height=4.7cm]{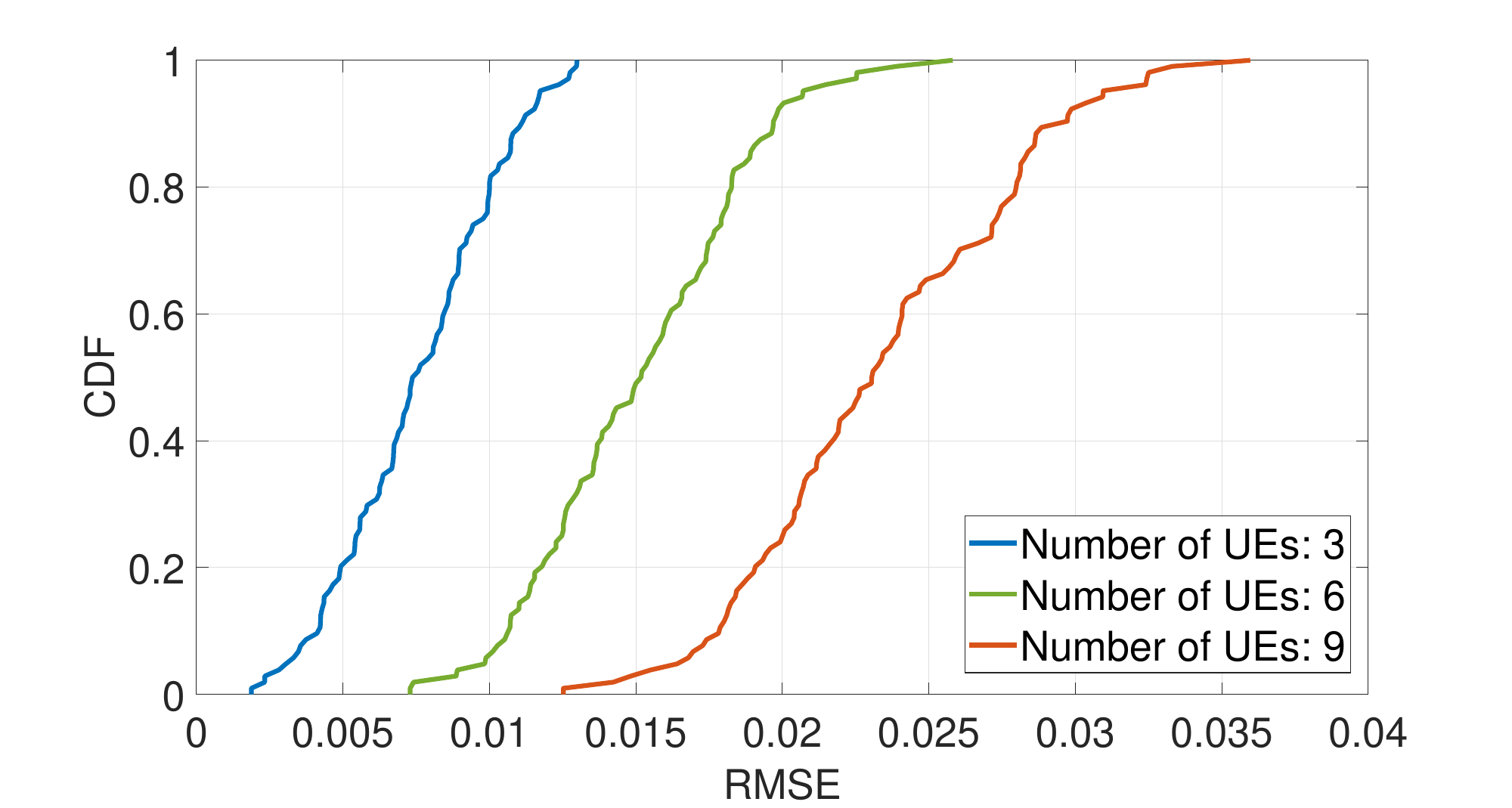}%\includegraphics
    \caption{CDF of the RMSE for angle estimation under different numbers of users ($N_t = N_r = 32$).}%
    \label{RMSE_CDF_ue_angle}
\end{figure}
%-------------------------------
\begin{figure}
    \centering
    \includegraphics[height=4.7cm]{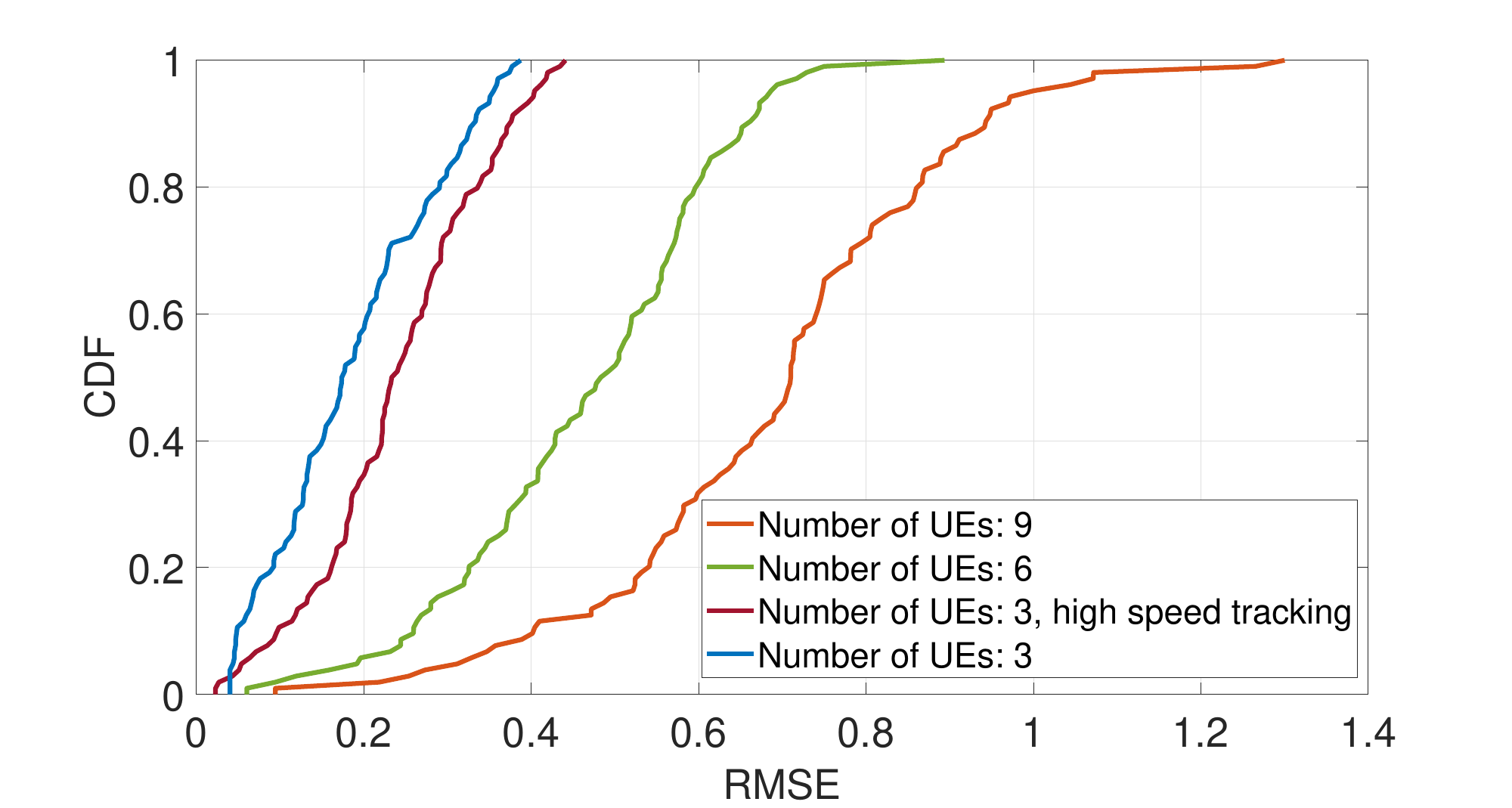}%\includegraphics[height= 7.500cm,width=7.8cm]{figures/CDF_ch_coefficient_1ue_diff_antenna_20_10_2025.pdf}
    \caption{CDF of the RMSE (m) for distance tracking under different numbers of users ($N_t = N_r = 32$).}%
    \label{RMSE_CDF}
\end{figure}

\begin{figure}
    \centering
    \includegraphics[height=4.7cm]{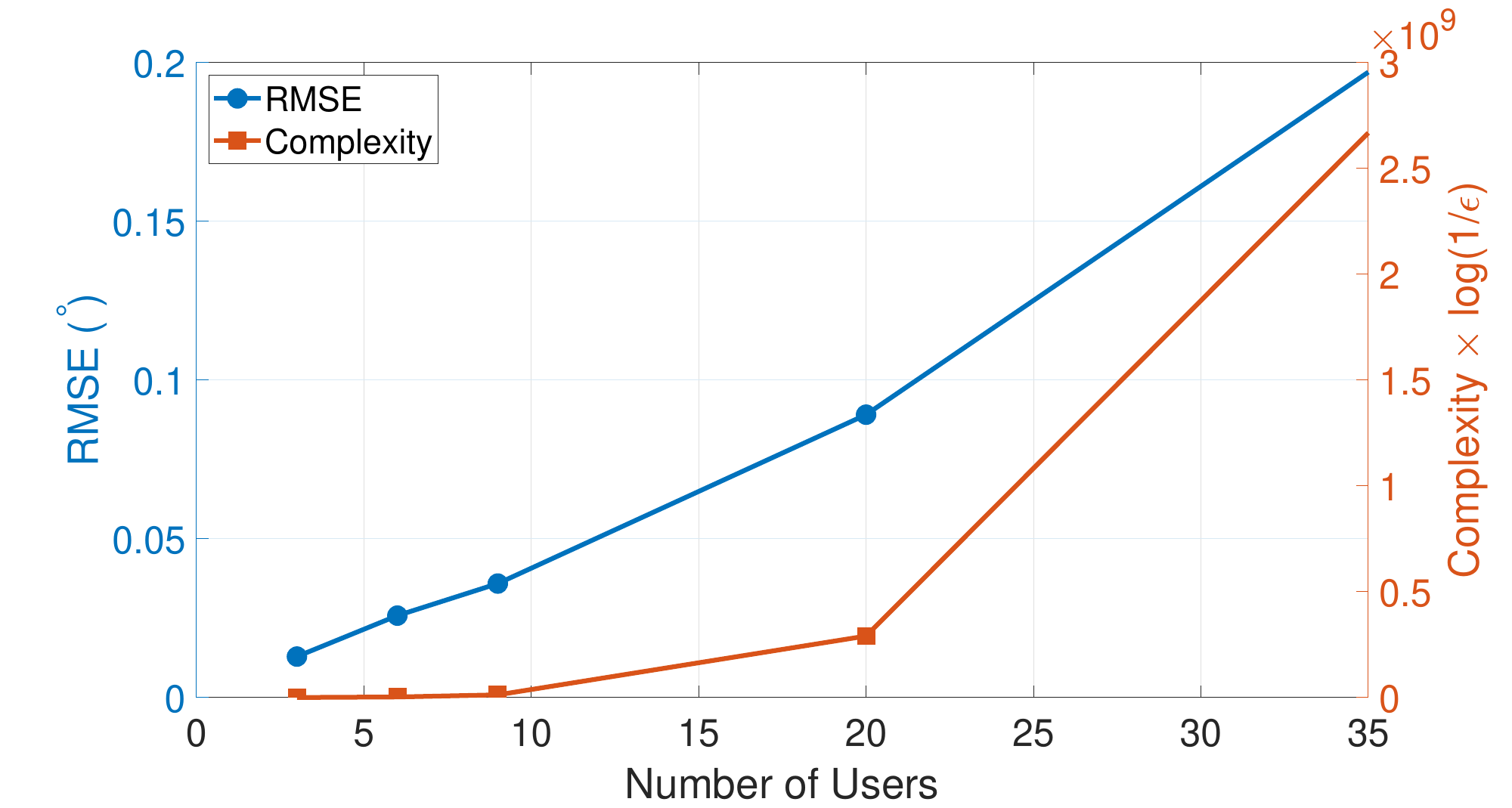}
    \caption{Computational complexity analysis and angle RMSE versus the number of users for one time slot.}%
    \label{comp_rmse_ues}
\end{figure}

\begin{figure}
    \centering
    \includegraphics[height=4.7cm]{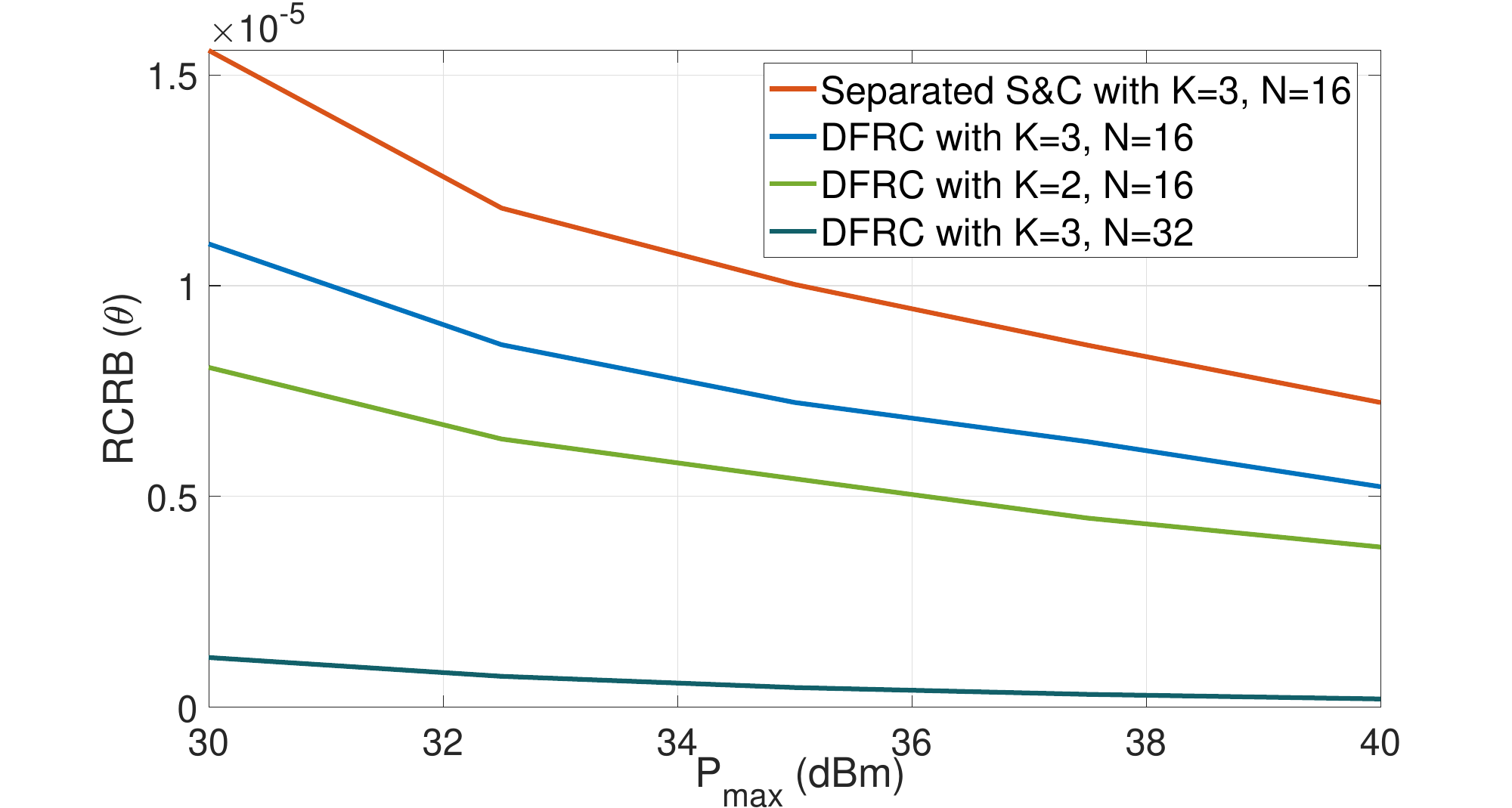}
    \caption{RCRB for different $P_{\text{max}}$ values.}%
    \label{RCRB_Pmax}
\end{figure}

\begin{figure}
    \centering
    \includegraphics[height=4.7cm]{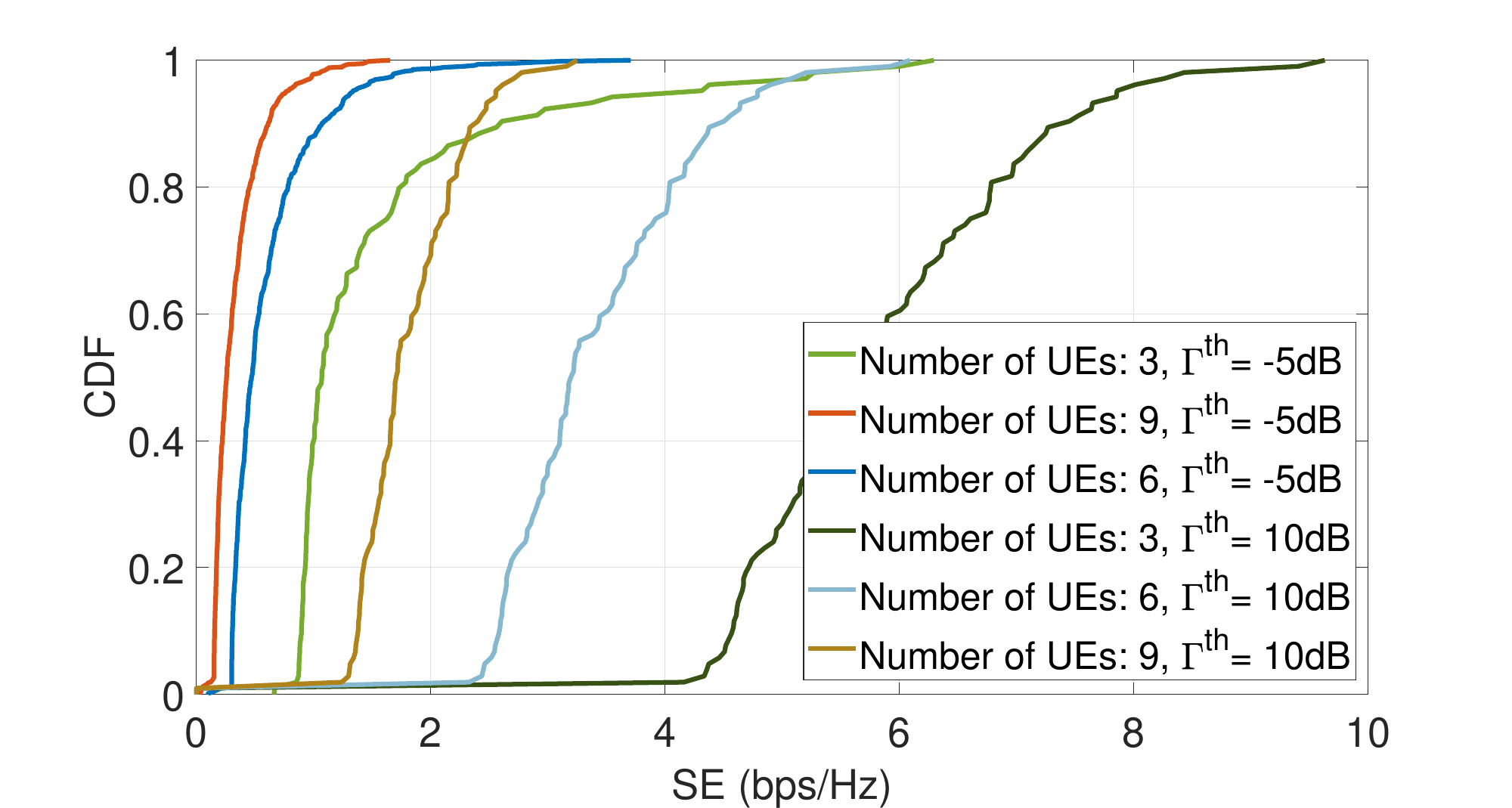}
    \caption{CDF of the achievable SE across all users and time slots.}%
    \label{cdf_rate_ues}
\end{figure}

The accuracy of angle estimation is evaluated using the \ac{CDF} of the \ac{RMSE} under different numbers of users, as illustrated in \Cref{RMSE_CDF_ue_angle}. The results show that increasing the number of users leads to higher estimation errors, with the degradation becoming more pronounced in denser scenarios (e.g., $K=9$) due to the amplified impact of sensing channel impairments.
Furthermore, the accuracy of distance estimation is evaluated under varying numbers of users, as shown in \Cref{RMSE_CDF}. The results indicate that performance degrades as the number of users increases. To examine the estimation accuracy under a more general setting, we consider a heterogeneous-velocity setting in which the velocities of the three users are gradually varied from their initial values by up to 30 km/h, rather than assuming fixed velocities throughout the tracking period. This captures the acceleration behavior of real vehicular traffic and the resulting heterogeneity in speed across users and over time. As shown in \Cref{RMSE_CDF}, an increase in the users' velocities leads to a higher range estimation error, indicating a degradation in estimation accuracy as the users move faster.

 To further evaluate the network under a relatively dense user deployment, \Cref{comp_rmse_ues} investigates the computational complexity and angle RMSE in the first time slot for different numbers of users. A total of 35 users are considered in the simulation, with their initial locations randomly and uniformly distributed in $x \in [0,70]$ m. This configuration provides a higher user load compared with the baseline setting and enables the scalability of the proposed approach to be evaluated under a relatively dense deployment. As shown in \Cref{comp_rmse_ues}, increasing the number of users results in higher angle RMSE and, more significantly, a substantial increase in computational complexity. This sharp increase in complexity is due to its dependence on the number of users through the term $K N_t(3K+1)^3$, which scales approximately as $\mathcal{O}(K^4)$ for a fixed $N_t$.

%\textcolor{blue}{We remark that the nine vehicle trajectories are enclosed within an area of approximately $A = 398.6~\mathrm{m}^2$. Hence, the $(K=9)$ scenario corresponds to an average trajectory-based vehicle density of approximately $ 2.26 \times 10^{4}~\mathrm{vehicles/km^2}$. Generally, the number of vehicles that can be simultaneously supported depends on the required communication SINR/data-rate threshold, noise power, allowable sensing estimation error, available antenna and transmission resources, and computational latency. For larger traffic densities, vehicle scheduling or clustering can be employed to divide the vehicles into subsets that are jointly optimized within each resource allocation interval.}

\Cref{RCRB_Pmax} shows the optimized \ac{RCRB} for angle estimation in the proposed framework. For each power budget constraint, the total \ac{RCRB} of all users is obtained by averaging over the entire time horizon. The results in \Cref{RCRB_Pmax} show the \ac{RCRB} performance for different values of $P_\text{max}$. As expected, increasing the sensing power leads to improved estimation accuracy. This effect is particularly pronounced in more challenging scenarios, such as when the number of users is high and the number of antennas is limited, where sensing performance becomes more sensitive to power allocation. Moreover, the figure highlights that reliable multi-user sensing requires both sufficiently high sensing power and a larger number of antennas to achieve reasonable accuracy.
Additionally, we assess tracking using a separate radar sensing and communication structure as a second baseline. In this case, we obtain a higher \ac{RCRB} value than when jointly sensing and communicating.

\subsection{Communication performance evaluation}

We remark that the primary objective of the proposed
optimization framework is to minimize the worst-user \ac{CRB} rather than to maximize the
achievable data rate. The communication requirement enters only as a
minimum-rate constraint C1 in $\mathcal{P}1$ that must be satisfied
while the beamforming resources are primarily allocated to improve
target estimation accuracy. Consequently, the achievable \ac{SE}
reported in this section is a byproduct of this accuracy-oriented
design, and depends on several system parameters beyond the algorithm
itself, such as transmit power, noise power, the number of antennas, and the
per-user SINR threshold $\Gamma^{\text{th}}$. To illustrate this
dependency and demonstrate that the proposed framework is not
inherently rate-limited, we further evaluate a scenario with a
higher $\Gamma^{\text{th}}$, thereby forcing the beamforming design
to accommodate a higher \ac{QoS} demand.
Here, we evaluate the achievable data rate of the proposed algorithm.
We consider a system bandwidth of $BW=200$ MHz and report communication
performance in terms of \ac{SE}, in bit/s/Hz, to
provide a bandwidth-independent evaluation; for reference, an \ac{SE} of
$1$ bit/s/Hz corresponds to an absolute data rate of $200$ Mbps under
this bandwidth. As shown in \Cref{cdf_rate_ues}, the achievable \ac{SE}
decreases as the total transmit power of the \ac{DFRC}-\ac{BS} is
shared among a larger number of users, leading to reduced sensing
accuracy and lower data rate. We also investigate the influence of the
demanded SINR threshold, $\Gamma^{\text{th}}$, on the achievable
\ac{SE}. To this end, $\Gamma^{\text{th}}$ is varied from $-5$ dB to
$10$ dB, corresponding to a required \ac{SE} of
$C^{\text{th}} = \log_2\left(1+\Gamma^{\text{th}}\right)$. As shown in
the results, increasing $\Gamma^{\text{th}}$ leads to an increase in
the achieved \ac{SE}, indicating that a higher SINR requirement drives
the system to allocate its available resources more effectively to
satisfy the increased communication demand.

\begin{figure}
    \centering
    \includegraphics[height=4.7cm]{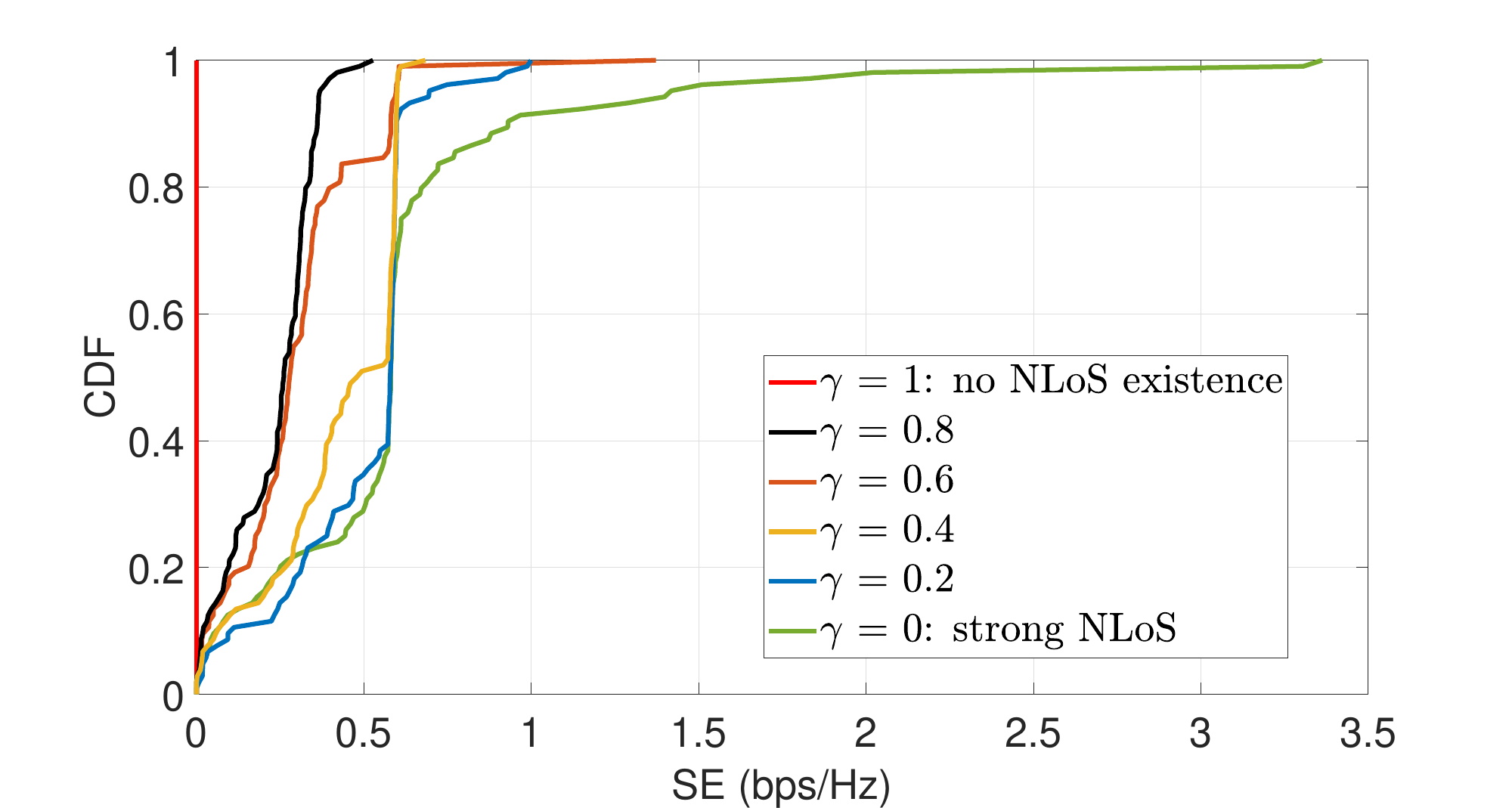}
    \caption{CDF of the achievable SE for the second UE without a LoS link ($K=3$).}%
    \label{cdf_rate}
\end{figure}

\begin{figure}
    \centering
\includegraphics[height=4.7cm]{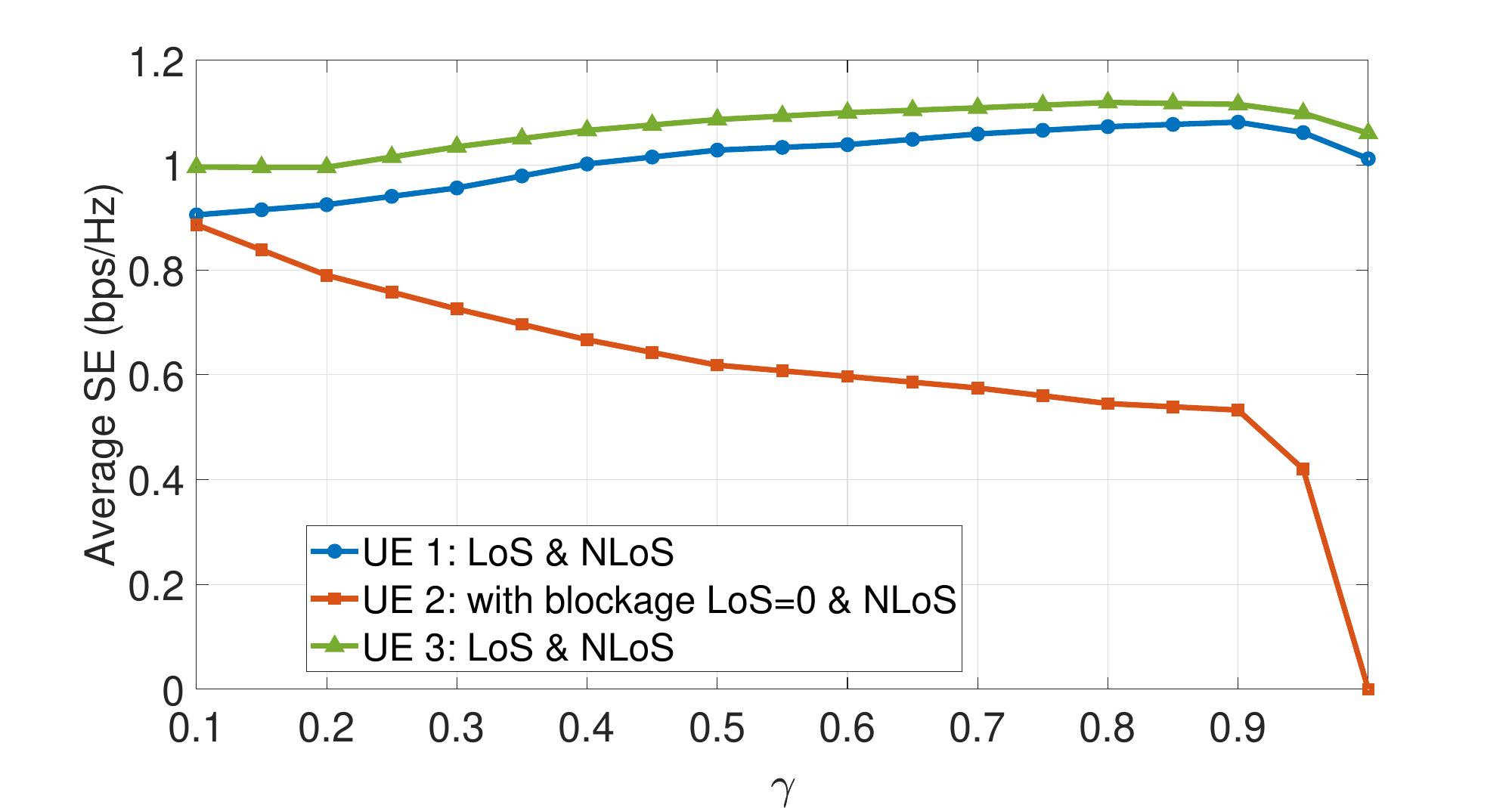}
    \caption{Average achievable SE of UEs for all time slots versus different NLoS coefficient contributions where the LoS link for UE 2 is totally blocked ($K=3$).}%
    \label{rate_3ue_gamma}
\end{figure}

 In order to focus on the communication channel modeling module and the impact of mobile scatterers on throughput, we consider a scenario in which one of the users lacks a \ac{LoS} path, i.e., $\mathbf{h}^{\text{LoS}}_{k,n}(t)=\mathbf{0}$ and analyze how the achievable data rate for that specific user changes based on the contributions of \ac{NLoS} links. For the scenario described, \Cref{cdf_rate} shows the \ac{CDF} of data rates for a specific blocked user across different time slots. The figure illustrates that when $\gamma = 1$, indicating the absence of \ac{NLoS} links, the blocked user cannot receive any data. However, as $\gamma$ decreases, the achievable data rate improves due to the enhanced contribution of scattering. Additionally, \Cref{rate_3ue_gamma} shows the average achievable data rates for a scenario with three UEs, where the \ac{LoS} path of the second UE is completely blocked. The results indicate that increasing $\gamma$, which corresponds to stronger \ac{LoS} links, improves the overall throughput. For the blocked user, however, weak \ac{NLoS} contributions lead to a decline in the data rate. Moreover, for the non-blocked users, there is a sweet spot where both \ac{LoS} and \ac{NLoS} links contribute significantly, occurring when $\gamma$ is between 0.7 and 0.9.

%%%%%%%%%%%%%%%%%%%%%%%%%%%%%%
%%%%%%%%%%%%%%%%%%%%%%%%%%%%%%%
%%%%%%%%%%%%%%%%%%%%%%%%%%%%%%

\subsection{Blockage detection effect} 
\begin{figure}
    \centering
    \includegraphics[width=0.48\textwidth]{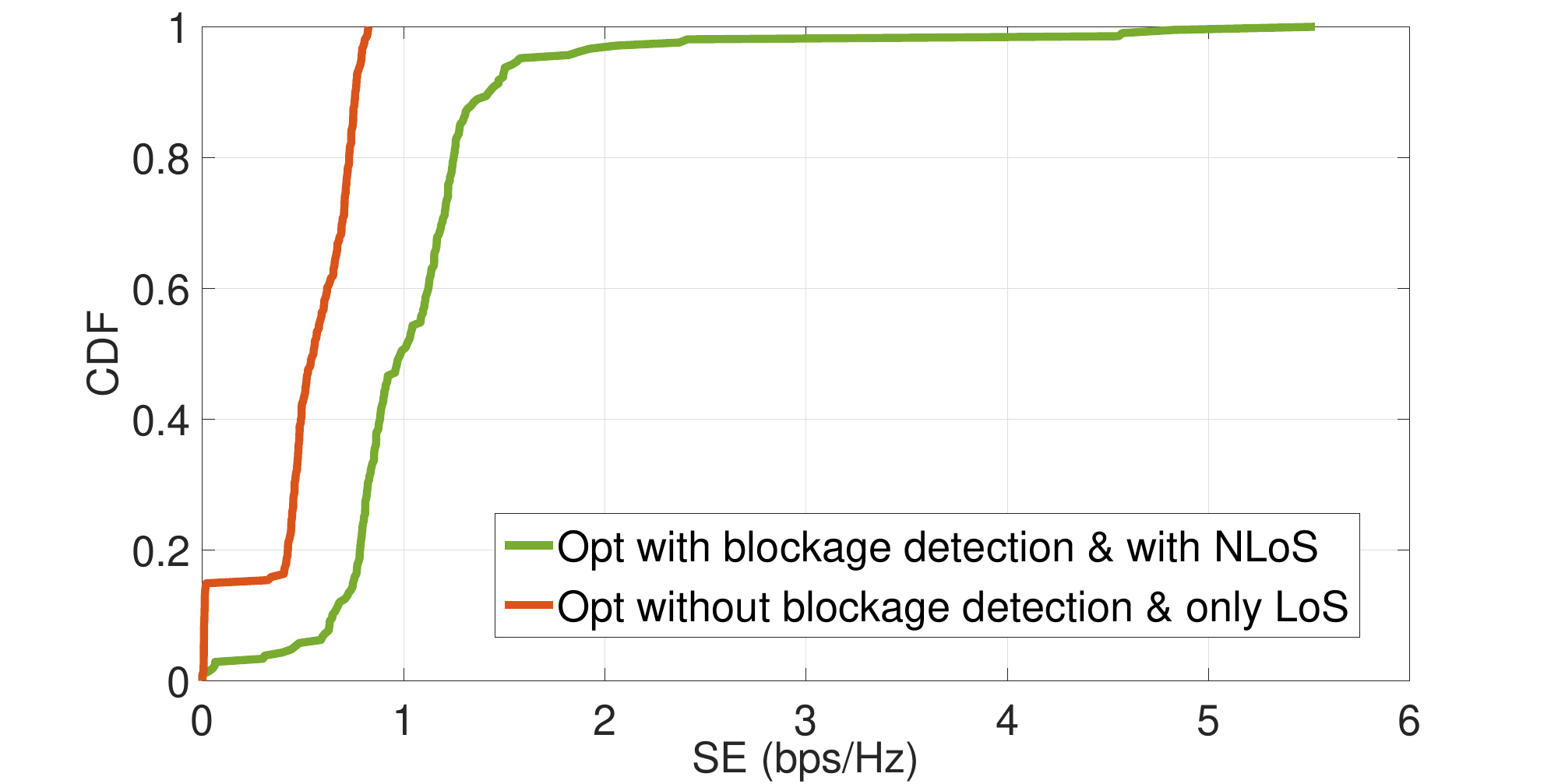}
    \caption{CDF of the achievable SE for all time slots ($K=3$).}%
    \label{cdf_rate_detection}
\end{figure}

In addition, we examine the reliability of our proposed framework under \ac{mmWave}-based network conditions, where signal propagation is highly susceptible to blockages. To this end, we consider a scenario in which one of the users encounters blockages during multiple time slots over the entire time horizon. The blockage is detected through the analysis of the received back-echo signal. Once a blockage is identified, the system promptly adapts the channel model to account for the \ac{LoS} link insufficiency. In this context, \Cref{cdf_rate_detection} illustrates the reliability of our proposed framework resulting from the detected \ac{NLoS} components. Specifically, the figure presents the \ac{CDF} of achievable data rates for all users across all time slots. The results indicate that, in the absence of \ac{NLoS} components, the system achieves lower throughput; in particular, when a user experiences a blockage, its data rate drops to zero. In contrast, by leveraging the modeled \ac{NLoS} channel based on tracked scatterers, the system can exploit these components during blocked periods, thereby maintaining connectivity for all users. Note that even with only a single user in the network, the presence of other non-detectable scatterers allows the user to maintain connectivity.

%%%%%%%%%%%%%%%%%%%%%%%%%%%%%%%%%%%%%%%%%%%%%%%%%%%%%%%%%%%%%%%%%%%%%%%%%%%%%%%%%%%%%%%%%%%%%%%%%%

\subsection{Robust design}
\begin{figure}
    \centering
\includegraphics[height=4.7cm]{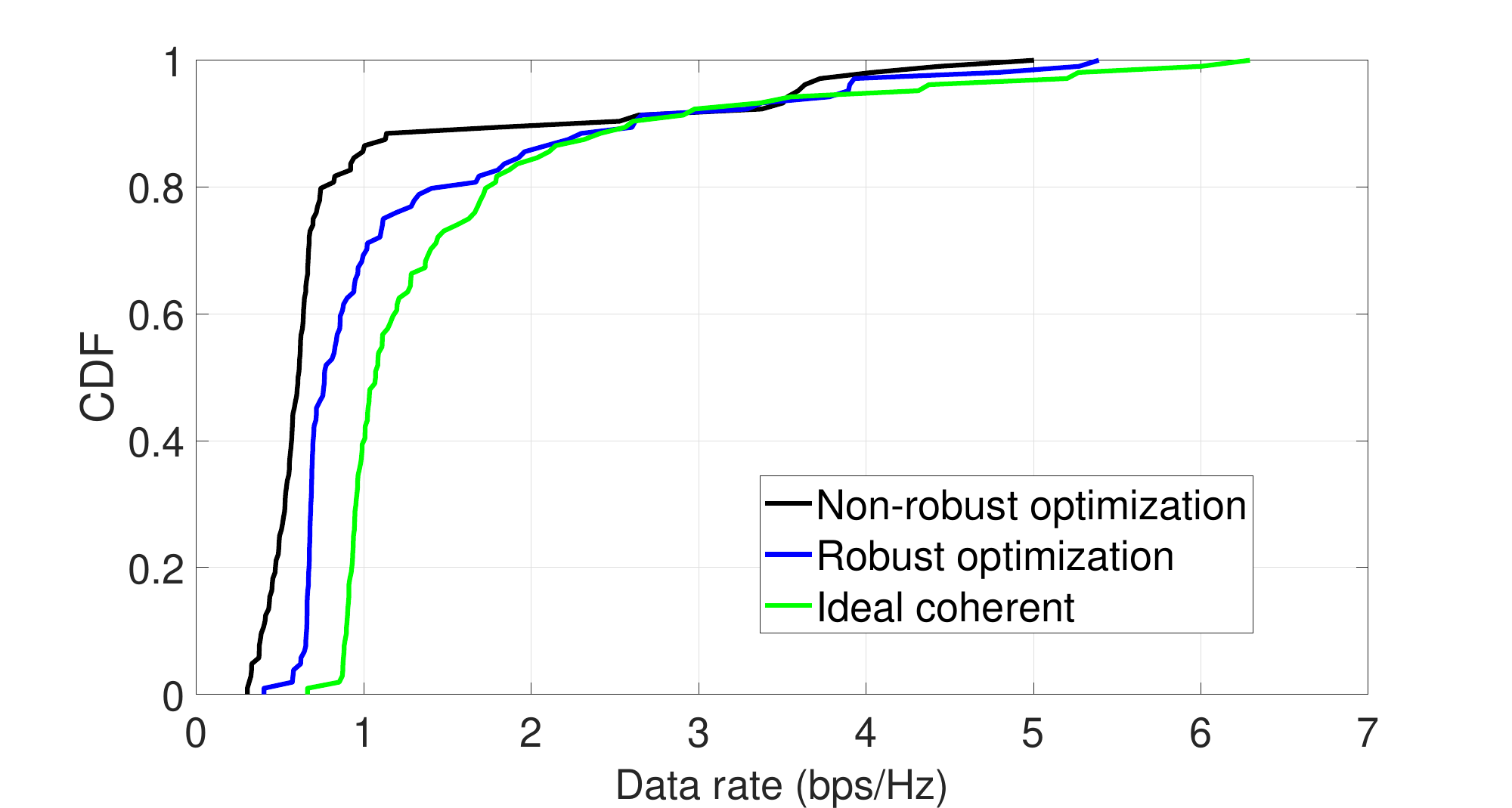}
    \caption{CDF of the achievable SE for predetermined absolute phase and random absolute phase ($K=3$).}%
    \label{cdf_rate_random_phase}
\end{figure}

Following the robust optimization problem $\mathcal{P}{5}$, we consider $U = 50$ samples for the \ac{SAA}. This choice strikes a balance between computational complexity and solution accuracy, ensuring that the approximation effectively captures the variability of the random phase parameter $\Phi_{k,n}$ while maintaining reliable convergence of the deterministic problem. In contrast, for problem $\mathcal{P}{1}$, $\Phi_{k,n}$ is assumed to have a fixed, known value, whereas in $\mathcal{P}_{5}$ it is treated as random. Accordingly, \Cref{cdf_rate_random_phase} illustrates the \ac{CDF} of the achievable data rates for all users across all time slots under both optimization schemes. The results indicate that, in the non-coherent case, accounting for phase uncertainties via robust beamforming yields higher achievable rates compared to a non-robust allocation scheme. Furthermore, coherent estimation with ideal phase knowledge outperforms non-coherent estimation, highlighting the performance benefits of accurate phase information.

%\section{Simulation Results}
%%%%%%%%%%%%%%%%%%%%%%%%%%%%%%%%%%%%%%%%%%%%%%%%%%%%%%%%%%%%%%

%-------------------------------------------------------------------------------
%-------------------------------------------------------------------------------

\section{Conclusion} \label{conclusion}

We proposed a sensing-assisted predictive beam tracking and localization framework for \ac{V2I} networks that combines geometry-based channel modeling, and \ac{EKF}-enabled channel prediction to tackle mobile \ac{mmWave} scenarios. By explicitly accounting for \ac{LoS} and \ac{NLoS} reflections from mobile scatterers, the framework accurately captures channel dynamics while reducing pilot overhead.
We further developed an optimization framework that dynamically maximizes sensing quality for the vehicle with the worst \ac{CRB} performance while accounting for network power constraints, resulting in improved robustness under varying vehicular positions and velocities. To enhance network reliability in practical deployments, a blockage detection mechanism was incorporated, exploiting backscattered echoes to maintain connectivity even when \ac{LoS} links are unavailable. In non-coherent scenarios, where absolute phase information is unavailable or impaired, a robust beamforming design was proposed to account for random phase noise, ensuring consistent system performance.
Simulation results demonstrate significant improvements in channel estimation accuracy, achievable data rate, and sensing reliability. The proposed approach effectively integrates accurate channel prediction, robust beamforming, and low-overhead tracking, providing a practical solution for high-speed \ac{ISAC}-enabled vehicular networks.
%%%%%%%%%%%%%%%%%%%%%%%%%%%%%%%%%%%%%%%%%%%%%%%%%%%%%%%%%%%%%%

\balance
\printbibliography

@book{boyd2004convex,
	author = {Boyd, Stephen and Vandenberghe, Lieven},
	title = {{Convex optimization}},
	doi = {10.1017/CBO9780511804441},
	isbn = {978-0-511-80444-1},
	publisher = {Cambridge University Press},
	year = {2004},
}

@article{du2023integrated,
	author = {Du, Zhen and Liu, Fan and Yuan, Weijie and Masouros, Christos and Yuan, Zenghui and Xia, Shuqiang},
	title = {{Integrated Sensing and Communications for V2I Networks: Dynamic Predictive Beamforming for Extended Vehicle Targets}},
	doi = {10.1109/TWC.2022.3219890},
	issn = {1558-2248},
	journal = {IEEE Transactions on Wireless Communications},
	month = Jun,
	number = {6},
	pages = {3612 -- 3627},
	publisher = {IEEE},
	volume = {22},
	year = {2023},
}

@article{el-ayach2014spatially,
	author = {El Ayach, Omar and Rajagopal, Sridhar and Abu-Surra, Shadi and Pi, Zhouyue and Heath, Robert W.},
	title = {{Spatially Sparse Precoding in Millimeter Wave MIMO Systems}},
	doi = {10.1109/TWC.2014.011714.130846},
	issn = {1558-2248},
	journal = {IEEE Transactions on Wireless Communications},
	number = {3},
	pages = {1499--1513},
	publisher = {IEEE},
	volume = {13},
	year = {2014},
}

@article{fernando2025sensing,
	author = {Pedraza, Fernando and Caire, Giuseppe},
	title = {{Sensing-Assisted Beam Tracking for mmWave V2I Communications With Analog, Hybrid, and Digital Antenna Architectures}},
	doi = {10.1109/TWC.2024.3495013},
	issn = {1558-2248},
	journal = {IEEE Transactions on Wireless Communications},
	month = Jan,
	number = {1},
	pages = {447 -- 461},
	publisher = {IEEE},
	volume = {24},
	year = {2025},
}

@article{gao2016channel,
	author = {Gao, Xiangyu and Dai, Linglong and Han, Shuhui and Chih-Lin, I and Heath, Robert W.},
	title = {{Channel estimation for millimeter-wave massive MIMO with hybrid precoding}},
	issn = {1941-0484},
	journal = {IEEE Journal of Selected Topics in Signal Processing},
	number = {3},
	pages = {435--446},
	publisher = {IEEE},
	volume = {10},
	year = {2016},
}

@article{guo2025integrated,
	author = {Guo, Hao and Wymeersch, Henk and Makki, Behrooz and Chen, Hui and Wu, Yibo and Durisi, Giuseppe and Keskin, Musa Furkan and Moghaddam, Mohammad H. and Madapatha, Charitha and Yu, Han and Hammarberg, Peter and Kim, Hyowon and Svensson, Tommy},
	title = {{Integrated Communication, Localization, and Sensing in 6G D-MIMO Networks}},
	doi = {10.1109/MWC.007.2400117},
	issn = {1558-0687},
	journal = {IEEE Wireless Communications},
	number = {2},
	pages = {214--221},
	publisher = {IEEE},
	volume = {32},
	year = {2025},
}

@article{heath2016overview,
	author = {Heath, Robert W. and Gonzalez-Prelcic, Nuria and Rangan, Sundeep and Roh, Wonil and Sayeed, Akbar M.},
	title = {{An Overview of Signal Processing Techniques for Millimeter Wave MIMO Systems}},
	doi = {10.1109/jstsp.2016.2523924},
	issn = {1941-0484},
	journal = {IEEE Journal of Selected Topics in Signal Processing},
	number = {3},
	pages = {436--453},
	publisher = {IEEE},
	volume = {10},
	year = {2016},
}

@article{khalili2024efficient,
	author = {Khalili, Ata and Rezaei, Atefeh and Xu, Dongfang and Dressler, Falko and Schober, Robert},
	title = {{Efficient UAV Hovering, Resource Allocation, and Trajectory Design for ISAC with Limited Backhaul Capacity}},
	doi = {10.1109/TWC.2024.3455370},
	issn = {1558-2248},
	journal = {IEEE Transactions on Wireless Communications},
	month = Nov,
	number = {11},
	pages = {17635--17650},
	publisher = {IEEE},
	volume = {23},
	year = {2024},
}

@article{li2025ekfbased,
	author = {Li, Ming and Dong, Fuwang and Liu, Tao and Liu, Fan},
	title = {{EKF-Based Beamforming Design for Joint Beam Tracking and Communication Systems}},
	doi = {10.1109/LWC.2025.3583503},
	issn = {2162-2337},
	journal = {IEEE Wireless Communications Letters},
	month = Jun,
	number = {6},
	pages = {2937--2941},
	publisher = {IEEE},
	volume = {14},
	year = {2025},
}

@article{liu2020joint,
	author = {Liu, Fan and Masouros, Christos and Petropulu, Athina P. and Griffiths, Hugh and Hanzo, Lajos},
	title = {{Joint Radar and Communication Design: Applications, State-of-the-Art, and the Road Ahead}},
	doi = {10.1109/tcomm.2020.2973976},
	issn = {1558-0857},
	journal = {IEEE Transactions on Communications},
	month = Jun,
	number = {6},
	pages = {3834--3862},
	publisher = {IEEE},
	volume = {68},
	year = {2020},
}

@article{liu2020radarassisted,
	author = {Liu, Fan and Yuan, Weijie and Masouros, Christos and Yuan, Jinhong},
	title = {{Radar-Assisted Predictive Beamforming for Vehicular Links: Communication Served by Sensing}},
	doi = {10.1109/TWC.2020.3015735},
	issn = {1558-2248},
	journal = {IEEE Transactions on Wireless Communications},
	month = Nov,
	number = {11},
	pages = {7704--7719},
	publisher = {IEEE},
	volume = {19},
	year = {2020},
}

@article{liu2025multipath,
	author = {Liu, Haotian and Wei, Zhiqing and Wang, Xiyang and Wu, Huici and Liu, Fan and Li, Xingwang and Feng, Zhiyong},
	title = {{Multipath Component-Enhanced Signal Processing for Integrated Sensing and Communication Systems}},
	doi = {10.1109/TCOMM.2025.3634233},
	issn = {1558-0857},
	journal = {IEEE Transactions on Communications},
	month = Mar,
	number = {3},
	pages = {1--1},
	publisher = {IEEE},
	volume = {73},
	year = {2025},
}

@phdthesis{moghaddam2023statistical,
	author = {Moghaddam, Mohammad H.},
	title = {{Statistical Analysis of Hardware Impairments in Communication Systems}},
	advisor = {Eriksson, Thomas},
	month = Apr,
	referee = {Eriksson, Thomas and Rezaei Aghdam, Sina},
	school = {Department of Electrical Engineering},
	type = {PhD Thesis},
	year = {2023},
}

@article{mohammadian2021rf,
	author = {Mohammadian, Amirhossein and Tellambura, Chintha},
	title = {{RF Impairments in Wireless Transceivers: Phase Noise, CFO, and IQ Imbalance -- A Survey}},
	doi = {10.1109/ACCESS.2021.3101845},
	issn = {2169-3536},
	journal = {IEEE Access},
	month = Jan,
	pages = {111718--111791},
	publisher = {IEEE},
	volume = {9},
	year = {2021},
}

@techreport{nguyen1989phaseambiguity,
	author = {Nguyen, Tien Manh},
	title = {{Phase-ambiguity resolution for QPSK modulation systems: Part 1: A review}},
	institution = {California Institute of Technology, NASA Jet Propulsion Laboratory},
	location = {Pasadena, CA},
	month = May,
	number = {NASA-CR-185385},
	type = {Contractor Report (CR)},
	year = {1989},
}

@article{raghavan2019statistical,
	author = {Raghavan, Vasanthan and Akhoondzadeh-Asl, Lida and Podshivalov, Vladimir and Hulten, Joakim and Tassoudji, Mohammad Ali and Koymen, Ozge Hizir and Sampath, Ashwin and Li, Junyi},
	title = {{Statistical Blockage Modeling and Robustness of Beamforming in Millimeter-Wave Systems}},
	doi = {10.1109/TMTT.2019.2899846},
	issn = {0018-9480},
	journal = {IEEE Transactions on Microwave Theory and Techniques},
	number = {7},
	pages = {3010--3024},
	publisher = {IEEE},
	volume = {67},
	year = {2019},
}

@book{scutari2018parallel,
	author = {Scutari, Gesualdo and Sun, Ying},
	title = {{Parallel and Distributed Successive Convex Approximation Methods for Big-Data Optimization}},
	doi = {10.1007/978-3-319-97142-1_3},
	edition = {1},
	isbn = {978-3-319-97142-1},
	publisher = {Springer, Cham.},
	series = {Lecture Notes in Mathematics},
	volume = {2224},
	year = {2018},
}

@article{snchez2021fading,
	author = {S{\'{a}}nchez, Jos{\'{e}} David Vega and Urquiza-Aguiar, Luis and Paredes, Martha Cecilia Paredes},
	title = {{Fading channel models for mm-Wave communications}},
	doi = {10.3390/electronics10070798},
	issn = {2079-9292},
	journal = {Electronics},
	number = {7},
	publisher = {MDPI},
	volume = {10},
	year = {2021},
}

@book{tse2005fundamentals,
	author = {Tse, David and Viswanath, Pramod},
	title = {{Fundamentals of Wireless Communication}},
	doi = {10.1017/CBO9780511807213},
	publisher = {Cambridge University Press},
	year = {2005},
}

@article{wang2024cramer,
	author = {Wang, Yiqiu and Tao, Meixia and Sun, Shu},
	title = {{Cram{\'{e}}r-Rao Bound Analysis and Beamforming Design for Integrated Sensing and Communication With Extended Targets}},
	doi = {10.1109/TWC.2024.3435864},
	issn = {1558-2248},
	journal = {IEEE Transactions on Wireless Communications},
	number = {11},
	pages = {15987--16000},
	publisher = {IEEE},
	volume = {23},
	year = {2024},
}

@article{yuan2021bayesian,
	author = {Yuan, Weijie and Liu, Fan and Masouros, Christos and Yuan, Jinhong and Ng, Derrick Wing Kwan and Gonz{\'{a}}lez-Prelcic, Nuria},
	title = {{Bayesian predictive beamforming for vehicular networks: A low-overhead joint radar-communication approach}},
	doi = {10.1109/TWC.2020.3033776},
	issn = {1558-2248},
	journal = {IEEE Transactions on Wireless Communications},
	number = {3},
	pages = {1442 -- 1456},
	publisher = {IEEE},
	volume = {20},
	year = {2021},
}

@inproceedings{zhao2023sensing-assisted,
	author = {Zhao, Yongkang and Xu, Xiaoli and Zeng, Yong and Liu, Fan},
	title = {{Sensing-Assisted Predictive Beamforming with NLoS Identification}},
	booktitle = {IEEE International Conference on Communications (ICC 2023)},
	address = {Rome, Italy},
	doi = {10.1109/ICC45041.2023.10278781},
	isbn = {978-1-5386-7463-5},
	issn = {1938-1883},
	month = May,
	pages = {6455--6460},
	publisher = {IEEE},
	year = {2023},
}

@article{xiao2016hierarchical,
    author = {Xiao, Zhenyu and He, Tong and Xia, Peng and Xia, Xiang-Gen},
    title = {{Hierarchical codebook design for beamforming training in millimeter-wave communication}},
    pages = {3380--3392},
    journal = {IEEE Transactions on Wireless Communications},
    issn = {1558-2248},
    publisher = {Institute of Electrical and Electronics Engineers (IEEE)},
    number = {5},
    volume = {15},
    year = {2016}
}

@article{hur2013millimeter,
    author = {Hur, Seungil and Kim, Taejoon and Lovell, David J. and {others}},
    title = {{Millimeter wave beamforming for wireless backhaul and access in small cell networks}},
    pages = {4391--4403},
    journal = {IEEE Transactions on Wireless Communications},
    issn = {1558-2248},
    publisher = {Institute of Electrical and Electronics Engineers (IEEE)},
    number = {10},
    volume = {61},
    year = {2013}
}

@article{alkhateeb2014channel,
    author = {Alkhateeb, Ahmed and Leus, Geert and Heath, Robert W.},
    title = {{Channel estimation and hybrid precoding for millimeter wave cellular systems}},
    pages = {831--846},
    journal = {IEEE Journal on Selected Areas in Communications},
    issn = {0733-8716},
    publisher = {Institute of Electrical and Electronics Engineers (IEEE)},
    month = {6},
    number = {6},
    volume = {32},
    year = {2014}
}

@article{rosadosanz2022adaptive,
    author = {Rosado‐Sanz, Javier and Jarabo‐Amores, Marı́a Pilar and De la Mata‐Moya, David and Rey‐Maestre, Nerea},
    title = {{Adaptive Beamforming Approaches to Improve Passive Radar Performance in Sea and Wind Farms’ Clutter}},
    pages = {6865},
    journal = {Sensors},
    issn = {1424-8220},
    publisher = {MDPI},
    number = {18},
    volume = {22},
    year = {2022}
}

@article{rojhani2024comprehensive,
    author = {Rojhani, Neda and Shaker, George},
    doi = {10.3390/rs16061016},
    title = {{Comprehensive Review: Effectiveness of MIMO and Beamforming Technologies in Detecting Low RCS UAVs}},
    journal = {Remote Sensing},
    publisher = {MDPI},
    number = {6},
    volume = {16},
    year = {2024}
}

@article{schmidt1986multiple,
    author = {Schmidt, Ralph},
    doi = {10.1109/tap.1986.1143830},
    title = {{Multiple emitter location and signal parameter estimation}},
    pages = {276--280},
    journal = {IEEE Transactions on Antennas and Propagation},
    issn = {0018-926X},
    publisher = {Institute of Electrical and Electronics Engineers (IEEE)},
    month = {3},
    number = {3},
    volume = {34},
    year = {1986}
}

@article{li2021joint,
    author = {Li, Xuan and Ma, Xiaochuan},
    title = {{Joint Doppler shift and time delay estimation by deconvolution of generalized matched filter}},
    pages = {29},
    journal = {EURASIP Journal on Advances in Signal Processing},
    issn = {1110-8657},
    publisher = {Hindawi},
    number = {29},
    volume = {2021},
    year = {2021}
}

@techreport{3gpp-38211,
    title = {{NR; Physical channels and modulation}},
    institution = {3rd Generation Partnership Project (3GPP)},
    location = {Sophia Antipolis, France},
    month = {6},
    number = {38.211 v19.0.0},
    type = {TS},
    year = {2025}
}

@inproceedings{lin2025integrated,
    author = {Lin, Xu and Du, Zhaopeng and Mei, Lin and Zhang, Ruoyu},
    doi = {10.1007/978-3-031-86203-8_20},
    title = {{Integrated Sensing and Communications: A Survey of Recent Progress Toward 5G-A and 6G}},
    address = {Harbin, China},
    booktitle = {Wireless and Satellite Systems. 14th EAI International Conference (WiSATS 2024)},
    month = {8},
    year = {2024}
}

@article{luo2010semidefinite,
    author = {Luo, Zhi-Quan and Ma, Wei-Kai and So, Anthony M-C and Ye, Yinyu and Zhang, Shuzhong},
    title = {{Semidefinite relaxation of quadratic optimization problems}},
    pages = {20--34},
    journal = {IEEE Signal Processing Magazine},
    issn = {1053-5888},
    publisher = {Institute of Electrical and Electronics Engineers (IEEE)},
    number = {3},
    volume = {27},
    year = {2010}
}

@article{chen2022generalized,
    author = {Chen, Li and Wang, Zhiqin and Du, Ying and Chen, Yunfei and Yu, F Richard},
    doi = {10.1109/JSAC.2022.3155515},
    title = {{Generalized Transceiver Beamforming for DFRC With MIMO Radar and MU-MIMO Communication}},
    pages = {1795--1808},
    journal = {IEEE Journal on Selected Areas in Communications},
    issn = {0733-8716},
    publisher = {Institute of Electrical and Electronics Engineers (IEEE)},
    month = {6},
    number = {6},
    volume = {40},
    year = {2022}
}

@article{liu2020transmit,
    author = {Liu, Xiang and Huang, Tianyao and Shlezinger, Nir and Liu, Yiming and Zhou, Jie and Eldar, Yonina C.},
    doi = {10.1109/TSP.2020.3004739},
    title = {{Joint Transmit Beamforming for Multiuser MIMO Communications and MIMO Radar}},
    pages = {3929--3944},
    journal = {IEEE Transactions on Signal Processing},
    issn = {1053-587X},
    publisher = {Institute of Electrical and Electronics Engineers (IEEE)},
    volume = {68},
    year = {2020}
}

@inproceedings{rezaei2026mobility-aware,
    author = {Rezaei, Atefeh and Carl, Mathis and Dressler, Falko},
    note = {to appear},
    title = {{Mobility-Aware Interference-Coupled ISAC for Energy-Efficient Multi-Vehicle Tracking}},
    publisher = {IEEE},
    address = {Paris, France},
    booktitle = {28th IEEE International Conference on Modeling, Analysis and Simulation of Wireless and Mobile Systems (MSWiM 2026)},
    month = {10},
    year = {2026},
   }

\end{document}